\documentclass{ws-jai}
\usepackage[flushleft]{threeparttable}
\usepackage{epsfig}
\usepackage{epstopdf}
\DeclareGraphicsExtensions{.pdf,.png,.jpg,.eps}
\begin{document}

\catchline{}{}{}{}{} % Publisher's Area please ignore

\markboth{Larry D'Addario and Douglas Wang}{An Integrated Circuit for Radio Astronomy Correlators Supporting Large Arrays of Antennas}

\title{An Integrated Circuit for Radio Astronomy Correlators\\Supporting Large Arrays of Antennas}
\author{Larry~R.~D'Addario$^\dagger$ and Douglas~Wang}
\address{Jet Propulsion Laboratory, California Institute of Technology, Pasadena, CA 91109, USA\\email:  ldaddario@jpl.nasa.gov}
\maketitle
\corres{$^\dagger$Corresponding author.}

\medskip

\begin{history}
\received{2015 October 5};
\revised{2016 January 11};
\accepted{2016 January 12};
Published 2016 March 18.
\end{history}

% Local definitions
\def\sysclk{{\it sysclk\/}}
\def\outclk{{\it outclk\/}}
\def\ceil{\mathop{\rm ceil}\nolimits}

\begin{abstract}
Radio telescopes that employ arrays of many antennas are in operation, and ever larger ones are being designed and proposed.  Signals from the antennas are combined by cross-correlation.  While the cost of most components of the telescope is proportional to the number of antennas $N$, the cost and power consumption of cross-correlation are proportional to $N^2$ and dominate at sufficiently large $N$.  Here we report the design of an integrated circuit (IC) that performs digital cross-correlations for arbitrarily many antennas in a power-efficient way.  It uses an intrinsically low-power architecture in which the movement of data between devices is minimized.  In a large system, each IC performs correlations for all pairs of antennas but for a portion of the telescope's bandwidth (the so-called ``FX" structure).  In our design, the correlations are performed in an array of 4096 complex multiply-accumulate (CMAC) units.  This is sufficient to perform all correlations in parallel for 64 signals ($N$=32 antennas with 2 opposite-polarization signals per antenna).  When $N$ is larger, the input data are buffered in an on-chip memory and the CMACs are re-used as many times as needed to compute all correlations.
The design has been synthesized and simulated so as to obtain accurate estimates of the IC's size and power consumption.  It is intended for fabrication in a 32 nm silicon-on-insulator process, where it will require less than 12\,mm$^2$ of silicon area and achieve an energy efficiency of 1.76 to 3.3\,pJ per CMAC operation, depending on the number of antennas.  Operation has been analyzed in detail up to $N=4096$.  The system-level energy efficiency, including board-level I/O, power supplies, and controls, is expected to be 5 to 7\,pJ per CMAC operation.  Existing correlators for the JVLA ($N=32$) and ALMA ($N=64$) telescopes achieve about 5000 pJ and 1000 pJ respectively using application-specific ICs in older technologies.  To our knowledge, the largest-$N$ existing correlator is LEDA at $N=256$; it uses GPUs built in 28 nm technology and achieves about 1000\,pJ.  Correlators being designed for the SKA telescopes ($N=128$ and $N=512$) using FPGAs in 16\,nm technology are predicted to achieve about 100\,pJ.
\end{abstract}


\section{Introduction}

As radio telescopes get larger, there is a need to provide digital signal processing electronics that are smaller and less power-hungry than would be implied by the extrapolation of existing designs.  This is especially true of correlation, which grows as $BN^2$ for $N$ antennas and processed bandwidth $B$.  The ALMA telescope in Chile \cite{alma}, with $N=64$ and $B=8\,$GHz, currently has the largest correlator by this measure, but much larger ones are planned.  Table 1 gives some properties of the correlators of existing and planned telescopes, including $N\approx 2000$.  This paper presents the design of an application-specific integrated circuit (ASIC or IC or chip) that enables construction of power-efficient correlators at large $N$.  The design provides considerable flexibility in that the IC can be used to construct correlators for a wide range of telescope sizes, from $N=32$ to essentially unlimited.

\begin{wstable}[h]\label{table:comparison}
\caption{Some Radio Telescope Correlators, Existing and Future}
\begin{tabular}{{@{}llccrrrrrrc@{}}}
\toprule
Telescope & Status & Technology & Year* & $w_i/2$** & $N$ & $B$, MHz & $P$, W & $2N^2B$, Hz & $P/2N^2B$, pJ & Notes \\
\colrule
VLA & obsolete & ASIC & 1975 & 1.5 & 27 & 200 & 50,000 & 2.92E+11 & 171,000 & a \\
JVLA & existing & ASIC 130 nm & 2005 & 4 & 32 & 8,000 & 70,000 & 1.64E+13 & 4,270 & b \\
ALMA & existing & ASIC 250 nm & 2002 & 3 & 64 & 8,000 & 65,000 & 6.55E+13 & 992 & c \\
LEDA & existing & GPU 28 nm & 2011 & 4 & 256 & 57.55 & 7,370 & 7.47E+12 & 977 & d \\
CHIME & existing & GPU 28 nm & 2013 & 4 & 128& 400 & 10,080 & 1.31E+13 & 769 & e \\
SKA1-low & proposed & FPGA 16 nm & 2017 & 8 & 512 & 300 & 11,600 & 1.57E+14 & 74 & f \\
SKA1-low &  & ASIC 32 nm & 2015 & 4 & 512 & 300 & 752 & 1.57E+14 & 4.8 & g \\
SKA1-mid & proposed & FPGA 16 nm & 2017 & 4 & 197 & 5,000 & 40,000 & 3.88E+14 & 103 & h \\
SKA1-mid &  & ASIC 32 nm & 2015 & 4 & 197 & 5,000 & 4,928 & 3.88E+14 & 12.7 & k \\
SKA2*** & planned & (TBD) & 2021 & TBD & 2000 & 5,000 & TBD & 4.00E+16 & (TBD) & m \\
\botrule
\end{tabular}
\begin{tablenotes}
\item All power estimates are for correlation only, at system level, including power supply loss but not including cooling.
\item * Design freeze year.
\item ** Input number quantization, real or imaginary part.  See section \ref{sec:parameters}.
\item *** Parameters are not yet well defined, so the values here are speculative.
\item[a] https://public.nrao.edu/gallery/radio-telescopes/image?id=300.
\item[b] \citet{mckinnen2010} and other sources; see also \citet{jvla}.
\item[c] 130 kW system \cite{almacorr}, approximately half for correlation.
\item[d] \citet{leda}.
\item[e] \citet{chime}.  Water cooled.  Data are for the CHIME Pathfinder; the full CHIME telescope will be larger.
\item[f] \citet{bunton}.  Preliminary power estimate, probably within a factor of two of the final value.
\item[g] This work.  Not equivalent to the currently proposed implementation due to smaller $w_i$.  Correlator ICs: $156\times$1.95\,W = 284\,W.  Support chips, power, control (est.): 468\,W.
\item[h] \citet{carlson}.  Preliminary power estimate, probably within a factor of two of the final value.
\item[k] This work.  Power is for $N=256$.  Correlator ICs: $1280\times 1.01$\,W = 1292\,W. Support chips, power, control (est.): 3636\,W.
\item[m] https://www.skatelescope.org/projecttimeline/.
\end{tablenotes}
\end{wstable}

\begin{figure}[h]
%\vskip -0.15in
\begin{center}
\includegraphics[height=1.9 in]{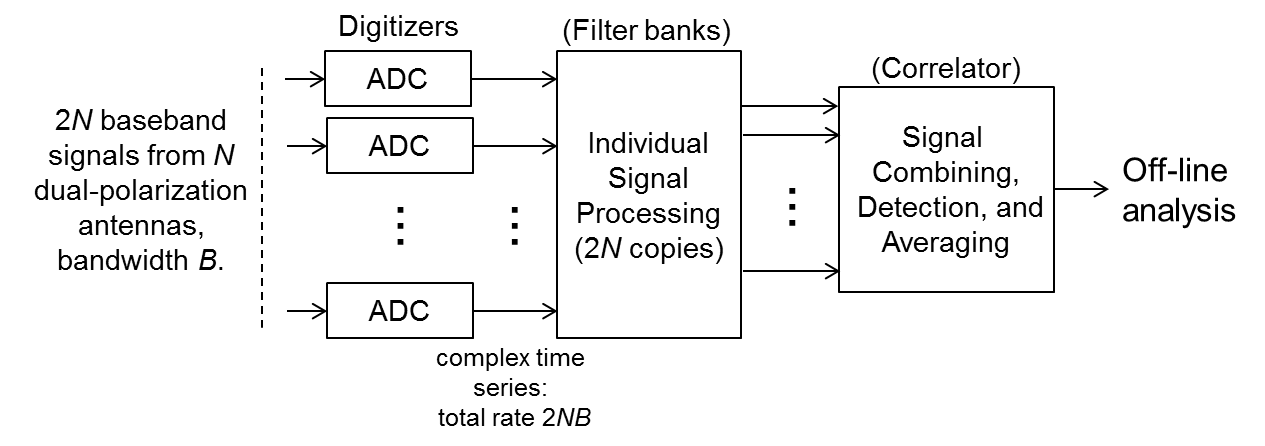}
\end{center}
\vskip -0.15in
\caption{Generic signal processing scheme for a multi-antenna telescope.  Each individual signal is often analyzed by a uniform filter bank and the signals are often combed by correlation, but other methods are sometimes used.}
\label{fig:generic}
\end{figure}

Figure \ref{fig:generic} is a simplified depiction of the signal processing required for a generic multi-antenna telescope.  
There are usually $2N$ input signals because each antenna collects radiation in two orthogonal polarizations.  Each signal is converted to baseband and digitized, forming a time series of complex numbers at rate $B$.  The processing is separated into operations performed on each signal separately and operations used to combine the signals.  Here we consider correlation as the combining method\footnote{Other combining methods, such as beamforming, are sometimes used, but correlation is often preferable because it allows forming an image by Fourier synthesis \cite{tms}.  Under special circumstances, combining via a spatial Fourier transform may be advantageous \cite{tegmark} \cite{morales}.}.  The rate of operations on individual signals is proportional to $BN$, so correlation dominates at sufficiently large $N$.  This paper concerns only correlation.

A correlator performs a complex multiply-accumulate (CMAC) operation on each sample of each pair of signals.  The total rate of CMAC operations is then $R = 2BN^2$ (including correlation of each signal with itself\footnote{The self-correlations produce real results and require only half the resources of a CMAC, so each physical multiply-accumulator can be configured to perform one cross-correlation or two self-correlations per cycle.  We exploit this in our design, and operations are counted accordingly.}).  A good figure-of-merit (FoM) for power efficiency is then $P/R$, where $P$ is the total power consumed.  Table 1 shows this FoM for each telescope (where only the power of the correlator proper, as shown in Fig.\ 1, has been included).  At the IC level (including all on-chip overhead and I/O), our design achieves 1.78\,pJ at $N=512$.  Table 1 includes estimates of the system-level FoM (including board-level I/O, power supplies, etc.) that would be achieved if correlators for the proposed SKA telescopes were based on this IC.  The design is most efficient when $N$ is a multiple of 64, so it is particularly inefficient for SKA1-mid at $N=197$.  Nevertheless, it would provide a substantial improvement over FPGA-based designs.

\begin{figure}[h]
%\vskip -0.15in
\begin{center}
\includegraphics[height=0.8 in]{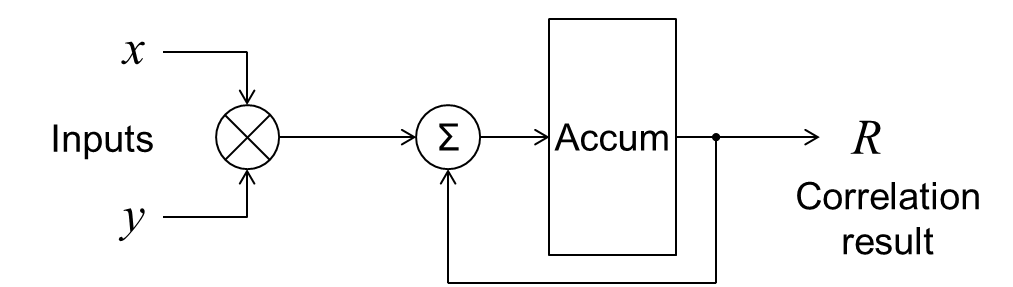}
\end{center}
\vskip -0.15in
\caption{Simplified block diagram of a complex multiply-accumulate (CMAC) unit.}
\label{fig:cmac1}
\end{figure}

\begin{figure}[h]
%\vskip -0.15in
\begin{center}
\includegraphics[height=0.75 in]{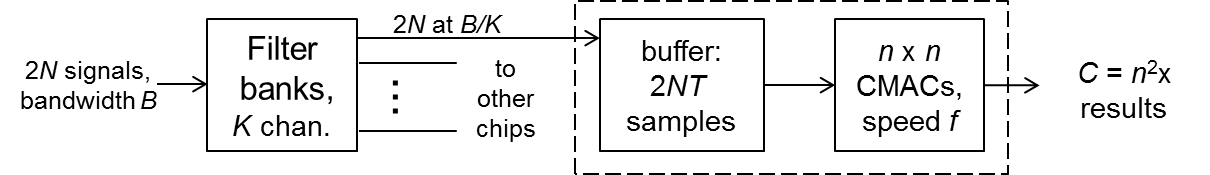}
\end{center}
\vskip -0.15in
\caption{Power-optimized correlator architecture.  The chip architecture is shown in the dashed box.  Each chip processes all $2N$ signals using an array of $n^2$ CMACs, but it does so for only a fraction of the bandwidth.  The number of cross-correlation products per chip is $2N^2$, which is typically $\gg n^2$, so the input data are buffered and the CMACs are re-used as many times as necessary.  A filter bank (not part of the correlator chip) breaks each input signal into frequency channels narrow enough so that one chip can process all signals.}
\label{fig:arch}
\end{figure}

Figure \ref{fig:cmac1} is a simplified block diagram of an individual CMAC unit.
It consists of a multiplier, an adder, and an accumulation register.  The product of each pair of samples is added into the accumulator for a fixed number of sample clocks $T$, then the accumulator content is copied to the output and the accumulator is cleared to begin another accumulation cycle or ``integration.''  Although each CMAC is straightforward, there are various architectures for organizing many CMACs to produce a large correlator.  The total CMAC rate is the same for all architectures, but the rates of other necessary operations (including memory writing and reading for temporary buffering of data and chip-level I/O) depend on the architecture.  A detailed study \cite{memo133} has shown that the chip-level architecture of Figure \ref{fig:arch} 
leads to the lowest system-level power consumption.  It achieves this by minimizing the rates of the auxiliary operations.  The architecture includes $n^2$ CMACs arranged as an $n\times n$ array and a memory that holds a block of $T$ input samples from all $2N$ signals.  When the total number of complex correlations to be computed ($2N^2$) is larger than the number of CMACs ($n^2$), the CMACs are re-used as many times as necessary by re-reading data from the buffer.   We call each computation of $n^2$ correlations one ``sub-integration'' (SI), and $x=2(N/n)^2$ SIs are needed to compute all the correlations of one full integration, where $x$ is called the CMAC re-use factor.

In practice this architecture requires that the buffer memory and CMAC array be on the same chip because of the high data rate between them.  The memory reading rate is $2N/n$ times higher than its writing rate, and this can easily lead to an impractical transfer rate if the CMAC array were on a separate chip.  (In our design, the transfer rate is up 352 Gb/s.)  Even if the transfer rate is practical, several orders of magnitude more energy is required to move a bit between chips than within a chip.  

Modern telescopes are too large for all correlations to be done in one IC, so the architecture must provide a way to partition the processing among many of them.  If each device has $n^2$ CMACs running at clock rate $f$, then $K = 2(N/n)^2 (B/f) = xB/f$ devices are needed.  One way to partition them is to have each device handle a subset of the signals, but then a copy of every signal sample must be delivered to at least $\sqrt{K}$ devices, leading to high total I/O rate and increased power consumption.  The favored architecture avoids this by having each device process all $2N$ signals (even if $N$ is large) but only a subset of the samples of each signal.  The samples could be decimated in time, but this would require another step in which the partial accumulations for the same signal pair are added across devices.  Instead, the samples are decimated in frequency.  Each signal is first processed by a uniform filter bank to separate it into at least $K$ independent channels of bandwidth $B/K$ (Fig.\ \ref{fig:arch}).  Each device then processes part of the bandwidth.  Because of the CMAC re-use, each device can handle input sample bandwidth $f/x = B/K$.  The filter bank could provide more than $K$ channels (but preferably an integer multiple of $K$), in which case each device can process several channels sequentially so that it still processes bandwidth $B/K$.  For further discussion of the reasons that this architecture is preferred and for descriptions of alternatives, see \citet{memo133}.

In this paper we describe a particular IC implementation that follows this architecture and takes additional steps to minimize power consumption.  Most existing synthesis telescopes whose correlators were considered large at the time of their design have used ASICs (including the original VLA, the expanded JVLA, the VLBA, and ALMA), but in those cases the ASIC was tailored to the particular requirements of one telescope.  Our design strives to be useful for a wide variety of future telescopes by supporting almost any number of antennas and any total bandwidth.

The design targets a 32 nm CMOS silicon-on-insulator process from IBM, 32SOI \cite{32soi}, where it will have a silicon area less than 12 mm$^2$.  The logic is specified in code written in Verilog, so in principle it could be ported to other processes, but our use of certain hard macros from IBM is not portable.  In particular, a high-density dynamic RAM is a critical component of the design.

\begin{figure}[h]
\begin{center}
\vskip -0.15in
\hbox{
\hskip -0.1 in
\includegraphics[width=0.495\textwidth, trim=0.57in 0in 0.6in 0in, clip]{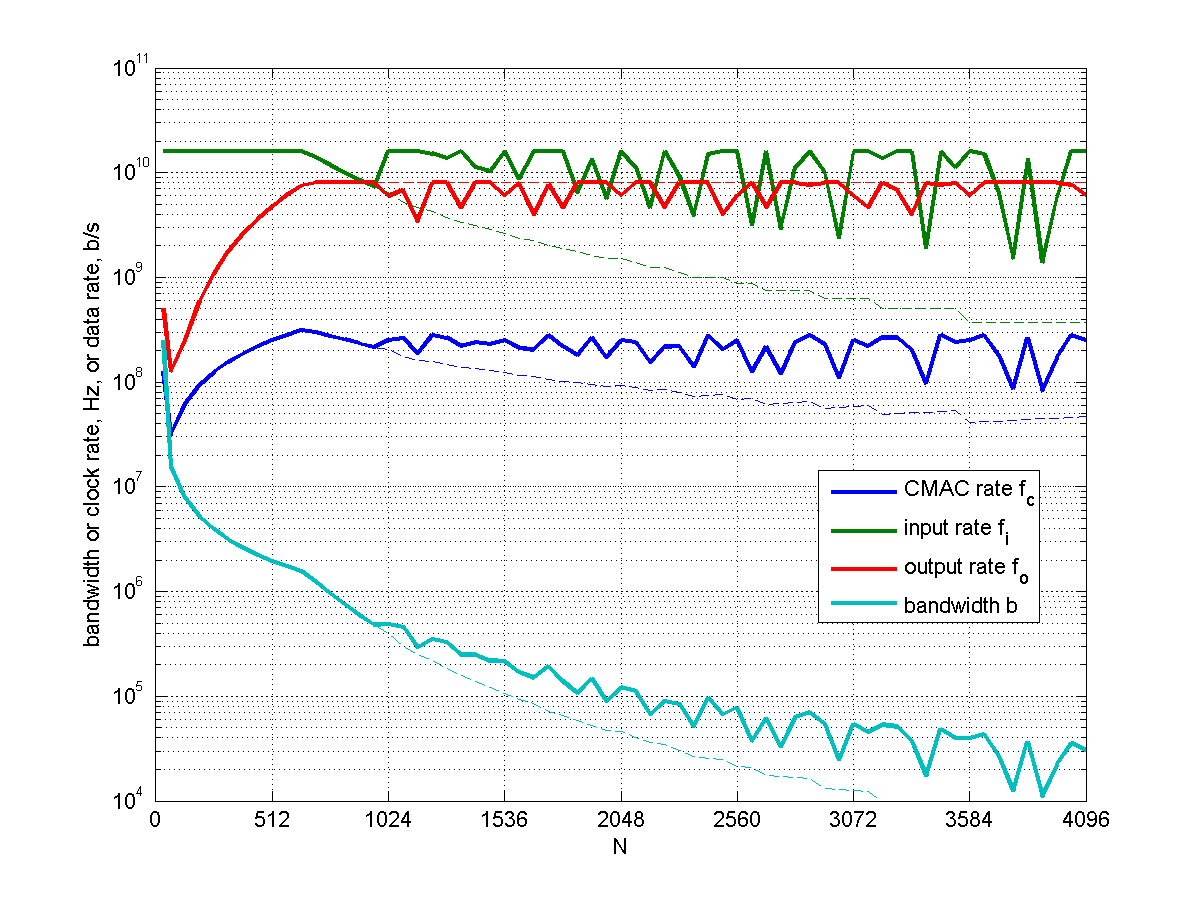}
\hskip -0.05in
\includegraphics[width=0.495\textwidth, trim=0.57in 0in 0.6in 0in, clip]{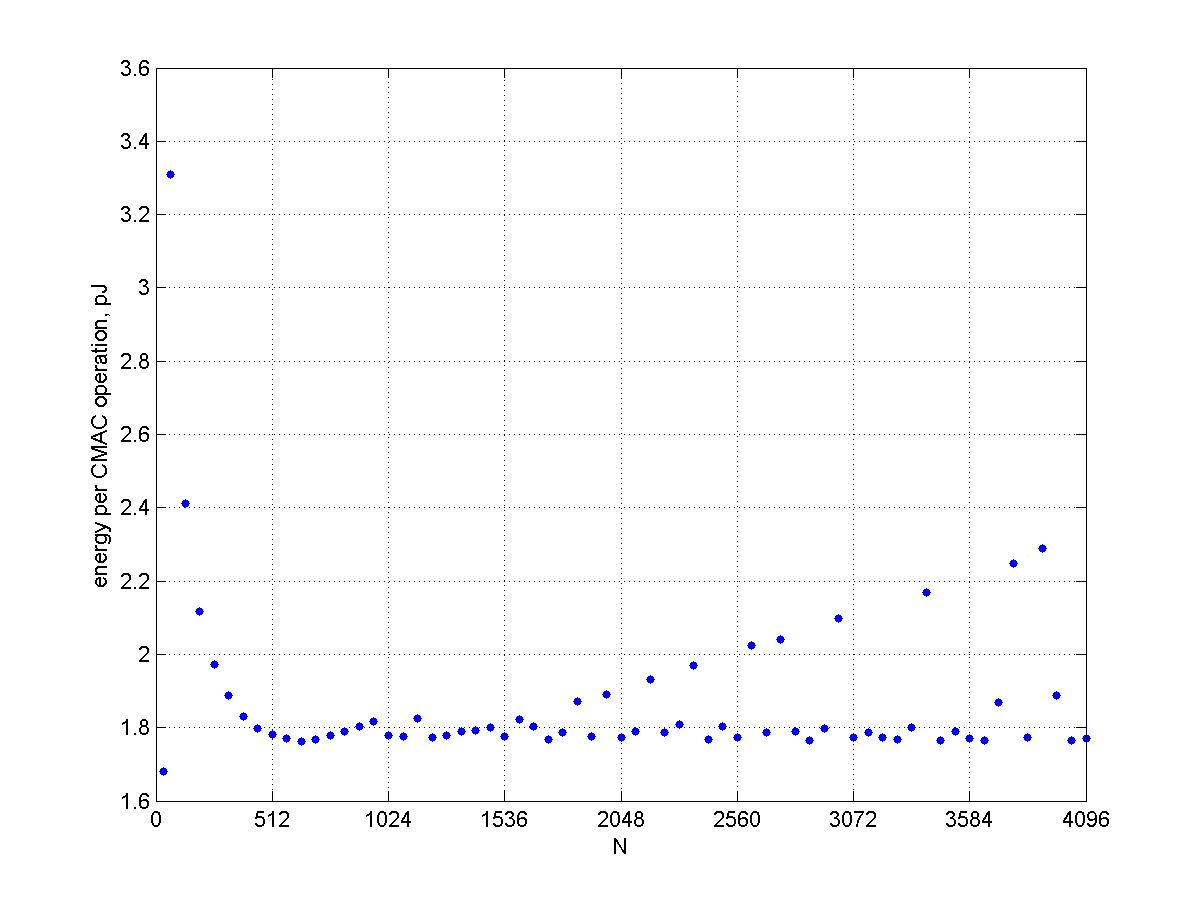}
}
\end{center}
\vskip -0.35in
\caption{Performance of the chip for $N=32$ and each multiple of 64.  Left:  bandwidth, I/O rates, and CMAC rate.   Right:  energy per CMAC operation (figure of merit for power).  For $N\ge1024$, the processing is broken into partial integrations (see section \ref{sec:largeN}).  Dashed lines show the performance that would be obtained if this were not done.  The energy per operation remains near its best value for most $N$, including arbitrarily large values.  The energy plot includes only the correlation chip power; additional power is used by auxillary devices needed for integrating the chips into a system (see section \ref{sec:integration}).}
\label{fig:performance}
\end{figure}

Figure \ref{fig:performance} summarizes the IC's performance over a wide range of $N$.  These results are discussed in more detail in section \ref{sec:PvsN}.  Results are computed for $N=32$ and each multiple of 64 up to 4096.  Except for $N=32$, which is a special case, the best performance occurs at $N=640$, where the device is utilizing its maximum input and output data bandwidths, but good performance is maintained over the full range.  In particular, the energy FoM achieves very nearly its minimum value of 1.76 pJ at many values of $N$ out to arbitrarily large values.  There are some unfavorable values of $N$ where the device cannot be used as efficiently, such as when $N/64$ is a prime number, but even then the performance remains good.  For an ideal device (where the processing rate remains constant and there is negligible overhead), the bandwidth decreases as $1/N^2$.  Fig.\ \ref{fig:performance} shows that the upper envelope of bandwidth vs.\ $N$ approaches this limit.

The remainder of this paper consists of an overview of the design (Section 2),  design details (3), synthesis and simulation results (4), performance as a function of $N$ (5), system integration (6), and conclusions (7).

\section{Design Overview}  %sec II

\subsection{Selection of fixed parameters}\label{sec:parameters}  %II.A

\begin{figure}[h]
%\vskip -0.15in
\begin{center}
\includegraphics[height=3 in]{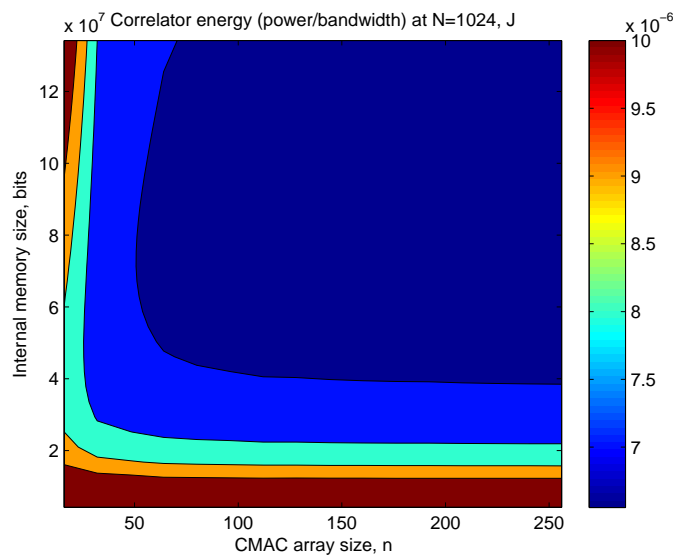}
\end{center}
\vskip -0.22in
\caption{Calculated energy per unit bandwidth vs.\ size of CMAC array and size of on-chip memory.  This result is based on a simple model that uses fixed values of the energies for each CMAC operation, for each memory operation, and for each I/O bit transferred, along with fixed values of the input and output word sizes.    Total I/O rates are unconstrained. The model parameters are approximately those expected for the IC described in this paper.}
\label{fig:energy1024}
\end{figure}

To create a specific design within the chosen architecture, it is necessary to select the number of CMACs $n^2$, the memory size, and the maximum input and output rates.  In general, more of everything is better, but there are practical limits to chip size and chip power dissipation, and compromises are needed to control cost.  If the IC were being designed for a fixed number of antennas $N$, more optimization would possible.  For example, Figure \ref{fig:energy1024} shows a model of power consumption per unit bandwidth vs. memory size $M$ and number of CMACs $n^2$ when $N=1024$.  
Although power decreases monotonically for increasing $M$ and $n$, the improvement is slow beyond $M=50$ Mb (at 8b per signal sample) and $n=60$.  For larger $N$, it is better to have more memory, even at the cost of fewer CMACs.  However, it was the objective of our design to support a wide range of $N$.  That is, we wanted the IC to be usable in multiple future telescopes whose sizes are not known in advance.  We chose $n=64$ and $M=64$ Mib, largely because this leads to a chip size that can be fabricated in an available process at an acceptable cost and yield.  It is shown in section \ref{sec:PvsN} that the resulting design is useful from $N=32$ to $N>4000$.

It is also necessary to choose the digital representations of the input and output data.  Each is a complex number, and we represent them as two twos-complement numbers with half of the bits used for the real and imaginary parts.  We choose $w_i=8$\,b for the input sample (4\,b real and 4\,b imaginary) and $w_o=32$\,b for the output result (16\,b and 16\,b).  

For the inputs, large words are expensive since the size and power consumption of the CMAC multipliers increases as $w_i^2$.  Conversely, large words provide smaller quantization noise and larger dynamic range.  With 4\,b numbers, quantization adds 1.2\% to the noise in measurements of cross-power between Gaussian-distributed random processes at the optimum signal levels \cite[Table 8.2, page 276]{tms}, and most correlator designs have recognized the diminishing returns of finer quantization%
\footnote{Even 1\,b (2 level) quantization is possible, and is used in some spectrometers \cite{vanvleck}.  VLBI has traditonally used 1\,b and 2\,b (4-level) quantization.  The original VLA correlator used 3 levels  or ``1.5 bits'' \cite{lrd1984}, the ALMA correlator uses 3\,b, and the JVLA and LEDA correlators use 4\,b.  See also Table \ref{table:comparison}.}  
Signal levels can be maintained near optimum by level controls in the filter banks or other up-stream circuitry.  Dynamic range is usually not a consideration in radio astronomy, since the signals are Gaussian-distributed processes with known variance.  An exception is when a signal includes unexpected interference.  Nearly all interference is human-generated, and those signals almost always have narrow bandwidth ($<1$\%) compared with astronomical observations ($10$\% to near 50\% in modern telescopes).   This can be exploited in FX processing (Fig.\ \ref{fig:arch}) where each signal is partitioned by frequency; those channels that are contaminated with interference can be ignored.  This may require maintaining high dynamic range (wide data words) in the per-signal processing, including the filter banks, but not in the correlator.

For the output words, size must be sufficient to avoid overflow during accumulation.  In our design, the internal accumulators are 20\,b+20\,b and 21\,b wide for cross- and self-correlations, respectively.  The maximum number of accumulations per integration is such that this is sufficient, even if the signal variances are more than 9\ dB above optimum.  The accumulators are rounded to 16\,b before delivery to the outputs.  Analysis shows that the rounding error is far smaller than the uncertainty caused by the intrinsic noisiness of the signals \cite{lrd2015}.  (See additional discussion in section \ref{sec:cmac}.)

\subsection{Block diagram}

\begin{figure}[h]
\begin{center}
\includegraphics[height=2 in]{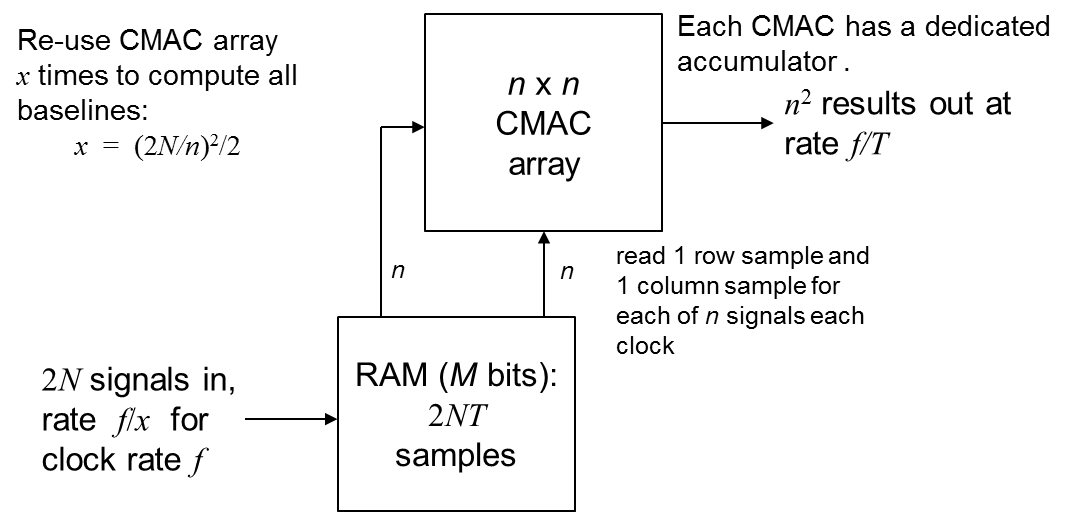}
\end{center}
\vskip -0.15in
\caption{High level block diagram of the correlator IC.}
\label{fig:block1}
\end{figure}

Figure \ref{fig:block1} is a high-level block diagram of the IC.  

Assume that the memory and CMAC array operate on the same clock.  On each clock cycle, one sample from each of $n$ row signals and $n$ column signals is read from the memory and presented to the CMAC array, which computes all $n^2$ cross-products and adds them to their respective accumulators.  After $T$ cycles, the accumulators are read out, completing one sub-integration.  Data are re-read from the memory for subsequent SIs until all correlations of the $2N$ signals have been computed, completing one integration.  Meanwhile, new data for the next full integration are being written to the memory from the input.  Because data in the memory are being re-used over $x$ SIs, the memory's write rate is slower than its read rate.  The data output rate depends on the integration length $T$, which is limited by the size of the memory.  

\begin{figure}[h]
\vskip -0.10in
\begin{center}
\includegraphics[height=3 in]{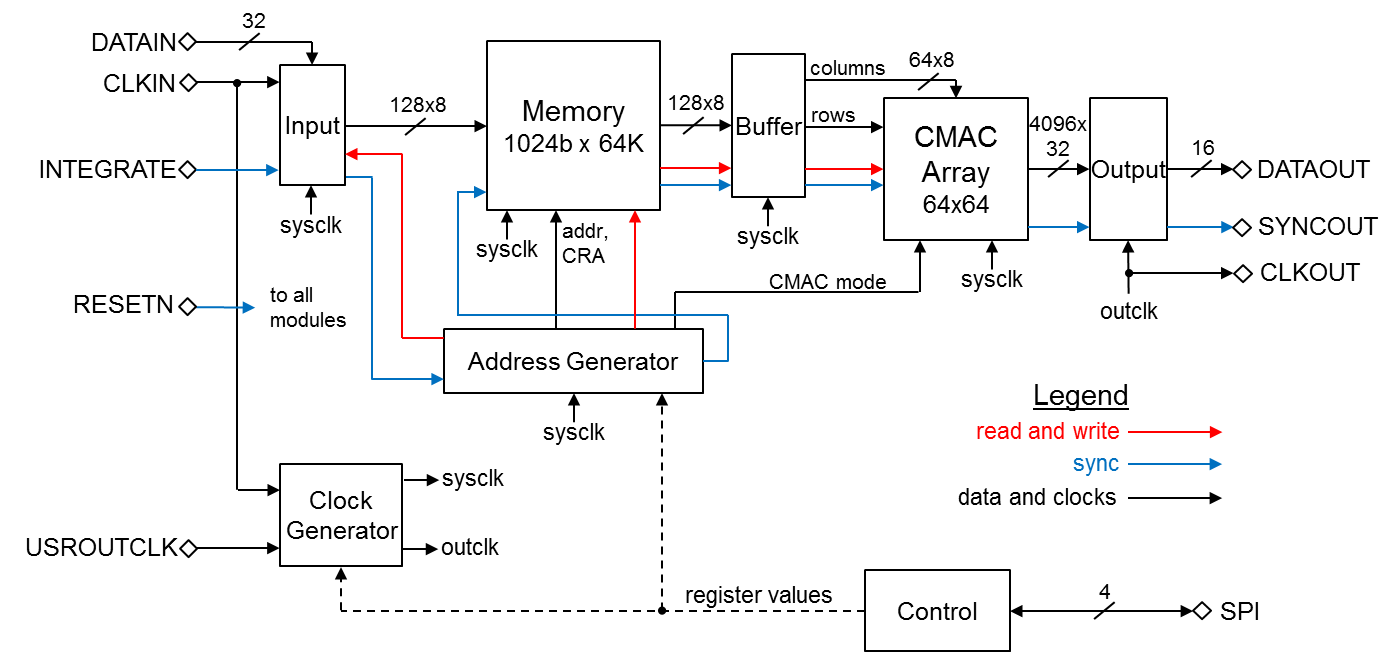}
\end{center}
\vskip -0.15in
\caption{Block diagram of the correlator IC, showing each of its major modules and all input and output ports.}
\label{fig:block2}
\end{figure}

Figure \ref{fig:block2} is a more detailed block diagram, showing individual modules of the implementation and all input/output signals.  
Input data are received at DATAIN, which is a 32\,b bus synchronized to the rising edge of CLKIN.  On each clock, four 8\,b samples of different signals at the same sampling time are received.  INTEGRATE, also synchronized to CLKIN, marks the start of a new integration.  The input module collects 32 of these 32\,b input words in a buffer, producing a 1024\,b word that is delivered to the memory module for writing to the RAM.  Successive input words supply samples from other signals and other sampling times in a particular order until all $2NT$ samples of one integration have been received and written to the memory.  The address generator module determines the sequencing of write and read cycles and provides the appropriate addresses for each.  It also provides synchronization signals that organize the chip's computations into sub-integrations.  The address generator is logically the most complex part of the design, although it uses very little silicon area; it is further described in the next section.  On memory read cycles, the RAM delivers 1024\,b words to a small buffer which organizes the data into $64\times 8$-bit words corresponding to the rows and columns of the CMAC array (for details see section \ref{sec:addrseq}).  The CMAC array module performs the correlations as previously described.  At the end of an SI, 4096 complex results are available.  These are delivered sequentially by the output module to 16 parallel output pins synchronous with 8192 cycles of CLKOUT.  SYNCOUT is asserted during the first CLKOUT cycle of each SI.  
%The frequency of CLKOUT must be large enough to read out all CMACs before the next SI is finished.
%, but it is allowed to be faster.  {include this if 'Details' section is deleted}

Most of the modules are driven by an internal clock called \sysclk, synthesized in the clock generator module from CLKIN using a phase-locked loop (PLL).  Each cycle of \sysclk\ is either a memory read cycle, memory write cycle, or an idle cycle, as determined by the address generator module.  The frequency of \sysclk\ must be sufficient to keep up with the input data.
%, but it is allowed to be faster.  {include this if 'Details' section is deleted}
During read cycles only, the CMAC array computes the appropriate correlations; otherwise it is idle.  The clock generator can also generate {\it outclk\/} for use by the output module and as output signal CLKOUT.  Alternatively, the output clock can be supplied externally at input USROUTCLK, in which case that signal is passed through clock generator to {\it outclk\/}.

The control module contains a set of twelve 20\,b registers that can be written and read by the user over a 4-wire serial peripheral interface (SPI) bus \cite{spi}.  These registers contain data for programming the PLL to set the frequencies of \sysclk\ and \outclk, as well as data needed by the address generator to determine the address sequence, including the values of $N$ and $T$.

\subsection{Operation}

The ability to control the clock frequencies and the address generation parameters provides considerable flexibility, making the chip useful for a wide range of $N$.  The integration length $T$ is also determined by a user-controlled register.  The frequency-decimation (FX) architecture of Fig.~\ref{fig:arch} allows building a correlator of almost any total bandwidth.

\begin{figure}[h] 
%\vskip -0.15in
\begin{center}
\includegraphics[height=2 in]{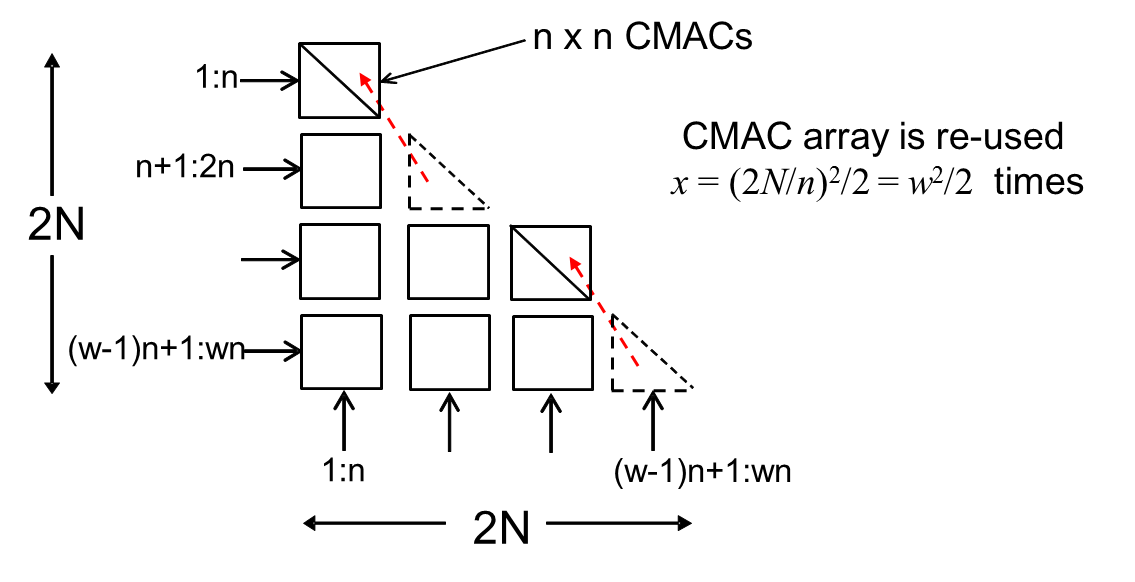}
\end{center}
\vskip -0.2in
\caption{CMAC re-use, illustrated for the case $2N/n=4$.  For our design, $n=64$ so $N=128$ in this example.}
\label{fig:reuse}
\end{figure}

Re-use of the CMACs follows a particular pattern, illustrated in Figure \ref{fig:reuse} for the case of $N=128$ (256 signals).  
In this figure, each square represents the same $n^2=4096$ CMACs performing one SI.  The SIs form a half-matrix with the $2N$ signals delivered to each row and column in groups of $n$.  There are $w=2N/n$ groups.  For the SIs along the diagonal, only half of the CMACs are needed to correlate the signals of a group with each other.  Therefore, the other half of the CMACs is used simultaneously for the correlation of another group.  During these SIs, the CMAC array operates in ``split mode," which is further described in section \ref{sec:cmac}.  In this way, only $w^2/2$ SIs are needed, not $w(w+1)/2$.  

It is desirable for $w$ to be an integer, and we will see later that it is best if $w$ is even.  To use the chip with a number of signals $2N$ that produces non-integer $w$, it is necessary to supply dummy signals so that the total number is a multiple of $n$.  Although the chip's throughput is only as large as it would be if the dummy signals contained valid values, its power consumption can be smaller.  If the dummy signals are always set to zero, no switching activity occurs within the CMAC array when they are being processed so no dynamic power is used.  We will show (sections \ref{sec:results} and \ref{sec:PvsN}) that the majority of the power is used by the CMAC array.

\begin{figure}[h]
%\vskip -0.15in
\begin{center}
\includegraphics[height=2.5 in]{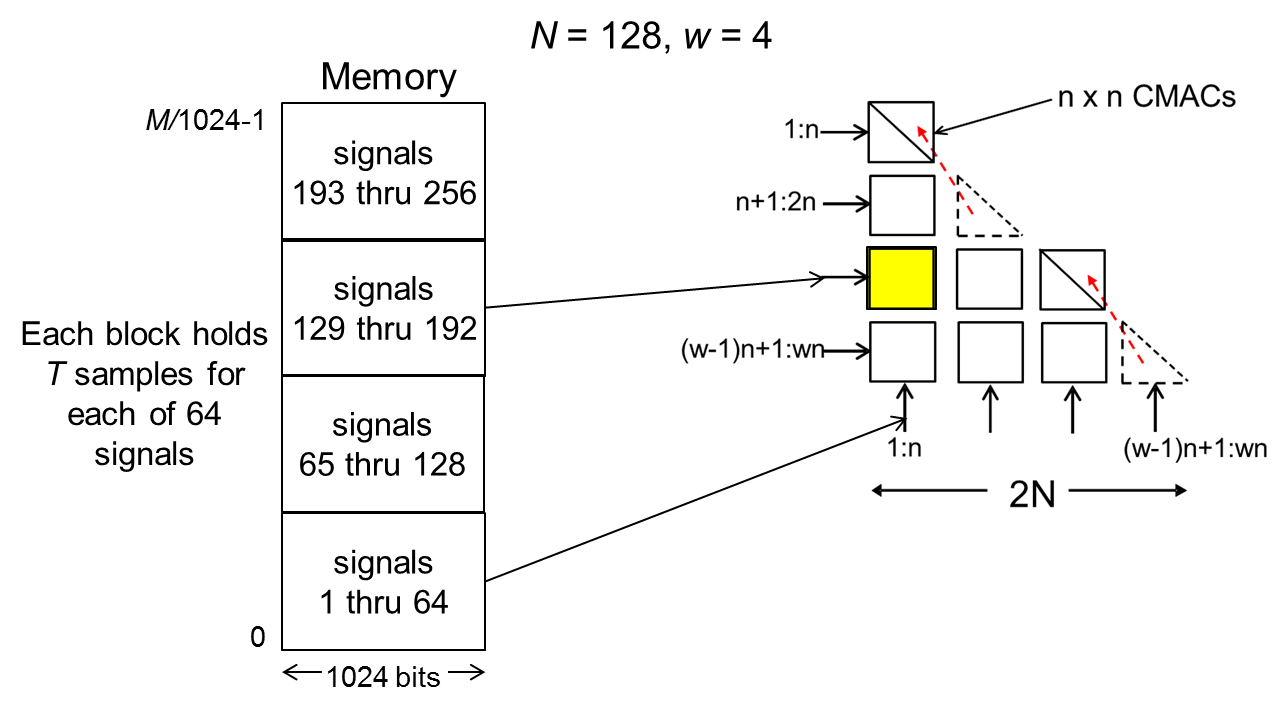}
\end{center}
\vskip -0.15in
\caption{Sub-integration pattern, illustrated for $2N/n=4$.}
\label{fig:subint}
\end{figure}

\begin{figure}[h]
%\vskip -0.15 in
\hbox{
\hskip -0.20 in
\includegraphics[width=0.55\textwidth]{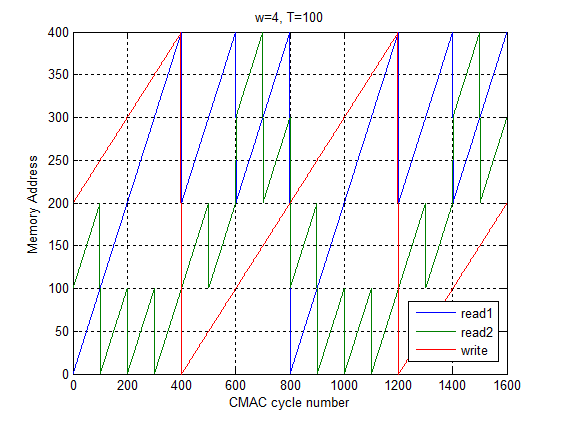}
\hskip -0.35 in
\includegraphics[width=0.55\textwidth]{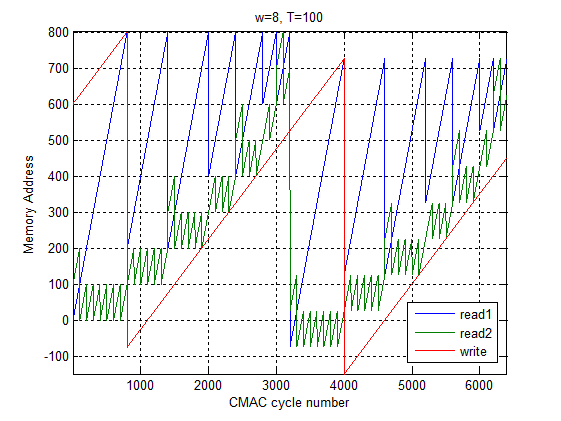}
}
%\epsfxsize=0.5\textwidth \epsfbox{addrplot-a.eps}
%\epsfxsize=0.5\textwidth \epsfbox{addrplot-b.eps}
%\special{psfile=addrplot-a.eps hoffset=-144}
%\special{psfile=addrplot-b.eps hoffset=144}
\vskip -0.15 in
\caption{Sequence of memory addresses plotted for $T=100$, showing the computation of two complete integrations, where read1 (blue) is the row data address, read2 (green) is the column data addreass, and write (red) is the write data address.  Left (a):  $w=4$.  Right (b):  $w=8$.  For $w>4$, some extra memory is needed to avoid overwriting data that has not yet been fully processed.  In plot (b), negative addresses should be interpreted modulo the complete memory size, which in this case must be at least 875 words.}
\label{fig:addrplot}
\end{figure}

The $w^2/2$ SIs can be computed in any order without changing the total computation effort, but the design uses a particular order so as to make efficient use of the memory.  This is illustrated in Figure \ref{fig:subint}, again for the case of $N=128$, $w=4$.  
Each of the four groups of signals is placed in a block of contiguous memory addresses, occupying $T/2$ words of 1024 bits.  The SIs are computed column-wise, beginning with the diagonal SI at the upper left of the diagram and proceeding downward.  The illustration shows computation of the third SI, involving signal groups 1-64 and 129-192.  Note that signal group 1-64 is needed for all four SIs of the first column, but when that column is complete that group is never needed again.  Therefore, as soon as the first column is complete the section of memory used for that group can be overwritten with data for the next full integration.  Similarly, when the second column is complete, the second memory block can be overwritten with new data.  Furthermore, data from the third and fourth groups is not needed for the current integration until SIs 3 and 4, so it can be written during SIs 1-4, as illustrated in Figure \ref{fig:addrplot}(a), which plots the memory addresses for reading column data, for reading row data, and for writing new data as a function of clock cycle number over two complete integrations.  
We see that complete overlap of reading and writing is achieved, with no need for double-buffering in order to have continuous operation.  Unfortunately this result cannot be extrapolated beyond $w=4$; to process more signals continuously, some additional memory is needed to buffer new data so that it does not overwrite data that are not yet fully processed.  For example, Figure \ref{fig:addrplot}(b) shows the situation for $w=8$, where 9.4\% additional memory is needed.  The additional memory requirement increases with $N$ but never exceeds 33\%, so full double-buffering (100\%) is not necessary.  The overlapping of reading and writing also means that the latency in delivering the results of an integration is less than the integration length; for example, at $w=8$ [Fig.~\ref{fig:addrplot}(b)] an integration is complete and all results are delivered to the ouptut 57 SIs (rather than 64 SIs) after the last data for that integration are delivered to the input.  In general, the latency is $w^2/2-w+1$ SIs.

This arrangement allows each block of memory to use a sequential set of addresses.  To compute one SI, we need only know the starting addresses of the row data (for reading), the column data (for reading), and the new data (for writing), each of which is simply incremented throughout the SI.  The starting addresses are given in control registers and are incremented automatically in the address generator logic and delivered to the memory in the proper sequence.  The user is required to update the three addresses via the SPI bus prior to the start of the next SI.  This is easily done with SPI clock rates that never exceed 15 MHz, although the chip is designed to support SPI rates of at least 25 MHz.

\subsection{Performance}

Performance is often limited by input or output bandwidth, as shown in Section \ref{sec:PvsN}.  Results given in this paper are based on achieving input and output clock frequencies of 500 MHz, or 16 Gb/s over 32 input pins and 8 Gb/s over 16 output pins.  SPICE simulations \cite{lrd2015c} show that this is easily achievable to or from a typical modern FPGA \cite{altera} via a wire-bond or flip-chip package and a 76-mm-long printed circuit board trace.  Considerably higher rates may be achievable if differential signaling is used for CLKIN and CLKOUT.  At I/O speeds of 500 Mb/s per pin, the best performance is achieved at $N=640$, where the bandwidth per chip is 1.5625 MHz and the total chip power is 2.26 W, giving a energy figure-of-merit of 1.76 pJ per CMAC operation.

\section{Design Details}\label{sec:details}

\subsection{Dynamic memory}

The selection of dynamic RAM (DRAM) for the on-chip memory was an important design choice.  The desired 64 Mib could lead to a large and expensive chip unless a high-density memory is used.   The DRAM available for the IBM 32SOI process achieves 2.8 times higher density than static RAM (SRAM) in the same process.  Even so, it occupies 65.5\% of the cell area in our design.  We considered other processes at about the same technology node and found that the available SRAMs had similar or lower density.  Few offered DRAM.  Therefore, use of SRAM was found to be unfeasible for the selected memory size.  Power was also a consideration; the selected DRAM uses only 14.3\% to 19.4\% of the chip's power, depending on $N$.  Use of a smaller memory was also considered, but it was found that this would require proportionally lower CMAC clock speed and bandwidth in order to maintain feasible output data rate.  Indeed, performance could be improved at large $N$ with more memory, so the selected size is already a compromise.

We restricted ourselves to processes for which multi-project wafer runs are available.  It is important to fabricate and test prototype devices before proceeding to production, and it is our judgment that funding for a dedicated prototype run would not be obtainable nor would it be justified.

Use of DRAM implies several design difficulties, including constraints on the sequence of read and write addresses and the need for periodic refreshing.  The DRAM has a concurrent refresh feature that allows refreshing to be overlapped with reading and writing, so in principle it is possible to avoid any refresh-only cycles.  However, the retention time is finite and this leads to a minimum clock speed of 200 MHz.  In addition, the DRAM version that we chose has a constraint that the same bank may not be addressed on successive clock cycles.  Each bank carries 2048 addresses, so our 65,536-address memory contains 32 banks.  On each clock cycle, it is necessary to supply the read or write address along with the refresh bank address (CRA in Fig.~\ref{fig:block2}); this is done by the address generator.  Every bank must get 256 refresh cycles within the retention time.  We have devised an algorithm that meets all constraints and guarantees refreshing all banks provided that the clock rate is at least 227 MHz.  If the total of the read rate and the write rate is at least this high, then no refresh-only cycles are needed.  Otherwise, $f_s$ is set to 227 MHz and the excess cycles are refresh-only.  During refresh-only cycles, the memory uses much less dynamic energy than during read or write cycles, and the CMACs use no dynamic energy because they are not clocked.

\subsection{Complex multiply-accumulator array}\label{sec:cmac}

\begin{figure}[h]
%\vskip -0.15in
\begin{center}
\includegraphics[height=1.5 in]{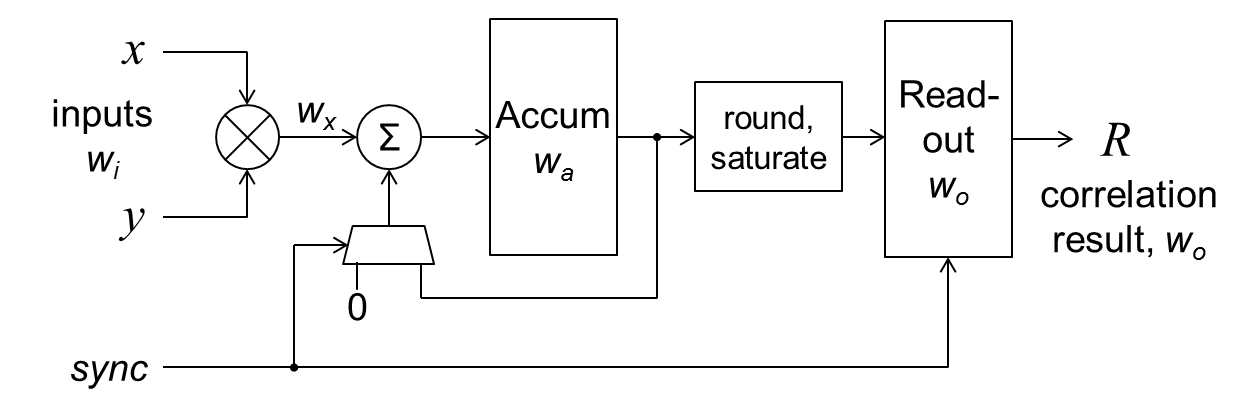}
\end{center}
\vskip -0.15in
\caption{Implementation of one CMAC module.  Inputs $x,y$, output $R$, and internal signals on the data path are complex numbers whose real and imaginary parts are each represented in twos complement with total size $w_i, w_x, w_a, w_o$ bits, as indicated.  On each clock, the product $xy$ is added to the accumulator.  Signal {\it sync\/} is asserted on the first clock of an integration.} 
\label{fig:cmac2}
\end{figure}

Our implementation of an individual CMAC is shown in Figure \ref{fig:cmac2}.  
The inputs $x$ and $y$ are each 4b+4b complex numbers ($w_i = 8$ in Fig.\ \ref{fig:cmac2}) in twos-complement form, with new values delivered at each clock.  To avoid any d.c.\ offset in the input data while retaining twos-complement representations, we assume that the range of the real and imaginary parts is [--7,+7]; that is, --8 is excluded. The complex multiplier is implemented with four real multipliers and two real adders%
\footnote{Complex multiplication can also be implemented with three real multipliers and five real adders \cite{mahdy}.  This can be advantageous when multiplication requires substantially more logic than addition, as happens with larger numbers but not with the 4\,b numbers used here.  Besides, the four-multiplier version facilitates implementation of the CMAC split mode, discussed later in this section.}, 
producing complex products that are no more than 8b+8b in twos complement ($w_x=16$).  Each product is added into a 20b+20b accumulator ($w_a=40$) for $T$ clock cycles.  On the next clock after that, signal {\it sync\/} is asserted; this causes the real and imaginary accumulator contents to be rounded to 16 bits each ($w_o=32$) and loaded into a holding register until they can be delivered to the output, and at the same time the product of the next pair of input samples is loaded directly to the accumulator, thus starting a new accumulation.

The accumulator size is chosen to avoid overflow in cross-correlation of Gaussian-distributed signals.  When correlating such signals, there is an optimum signal strength that minimizes quantization noise in the results; for 4\,b numbers, that optimum is a standard deviation of $\sigma_{\rm opt}=2.94$ on both the real and imaginary parts \cite[Table 8.2, page 276]{tms}.  It has been shown \cite{lrd2015} that even when the standard deviations of both signals are $3\sigma_{\rm opt}$ (9.5 dB above optimum) overflow does not occur at 20\,b until $T> 8,858$, even when the correlation coefficient is unity at the worst-case correlation phase.%
\footnote{In these calculations, overflow refers to the expected value of the accumulation, so the results are statistical.  Whereas the number of accumulation cycles is many thousand, the actual value is, with high probability, within a few percent of the expected value.}
  At $T=32,768$, overflow does not occur provided that the correlation coefficient is less than 0.27, also at $3\sigma_{\rm opt}$ and worst-case phase.  For our design, the memory size ensures that $T\le 32,768$ for $N > 64$.  High cross-correlation coefficients are rarely seen in radio astronomy (where $\ll .01$ is routine), but in situations where they can occur the signal level can be reduced upstream of the correlator.  If overflow does occur, the design ensures that it is detected.  A check for overflow is done on every clock, and if it ever occurs the final result is set to the maximum positive or negative value, as appropriate.

The rounding logic reduces the results from 20b+20b to 16b+16b by rounding off the 4 least-significant bits.  The mid-point is always rounded away from zero so that no bias is introduced\footnote{The cross-correlations have random phase, so the real and imaginary parts are symmetrically distributed about zero.  For the self-correlations, which are always positive, unbiased rounding is not possible with deterministic logic.  In practice, that bias is negligible because self-correlation values are large.}.  Rounding is acceptable to the extent that it does not cause loss of significance when the correlation coefficient is small.  Analysis shows \cite{lrd2015} that the quantization error due to rounding will be smaller than the intrinsic noise in the result provided that $T>31$ when the signal strengths are 9.5 dB below optimum, even if the true correlation coefficient is zero.    For our design, we are limited to $T\ge 512$ at all $N$, otherwise the output data rate would exceed the available output bandwidth.  This means that the rounding never introduces significant quantization noise.

When the CMAC array is computing sub-integrations on the diagonal of the SI half-matrix (Fig.~\ref{fig:reuse}), it must operate differently.  The $n\times n$ array then contains three versions of the CMAC cell:  CMACs below the diagonal are computing cross-correlations among the $n$ signals of one group (the ``row" signals); those above the diagonal are computing cross-correlations of another group (``column" signals); and those on the diagonal are computing two self-correlations at once, for one signal from each group.  The latter is possible because self-correlations accumulate the squared magnitude of the signal, and this can be done using only half the resources of a CMAC (two of the four real multipliers in the complex multiplier, and only the real part of the complex accumulator).  The other half of this CMAC is then used simultaneously to compute the self-correlation of a signal form the other group, accumulating its real result in the ``imaginary'' part of the accumulator.  This amounts to a re-organizing of the CMAC array when a diagonal SI is being computed.  We call this the ``split mode'' of the CMAC array, and it is controlled by a signal from the address generator called ``CMAC mode'' in Fig.~\ref{fig:block2}.

Self-correlation is equivalent to cross-correlation with a correlation coefficient of unity, so overflow can occur more easily.  On the other hand, the result is always positive so the accumulator can be considered unsigned; this allows $T$ to be twice as large before overflow occurs.  Nevertheless, to allow $T=32,768$ without overflow when the signal strength is 9.5 dB above optimum requires 21\,b.  Therefore, our implementation provides 21b+21b accumulators for the 64 CMACs on the diagonal, since they are the only ones that do self-correlations (and 20b+20b accumulators for the other 4032 CMACs).  In split mode only, those CMACs round off 5 least-significant bits rather than 4 so that the result is still $w_o = 16+16$ bits.

\subsection{Clocks}

To compute all the correlations of one integration, a total of $2N^2T$ CMAC operations is required, and since $n^2$ of them are done at once, this requires a total of $2(N/n)^2T = w^2T/2$ CMAC clocks.  At the same time, $2NT$ input data samples for the next integration must be received and written to the memory.  Each memory word carries $2n$ samples, so $(N/n)T$ memory write cycles are needed.  It follows that a memory write cycle is needed for every $w=2N/n$ CMAC cycles.  Since we also need one memory read cycle for each CMAC cycle, we need $w+1$ memory cycles (read+write) for each $w$ CMAC cycles.  Since 32b$/w_i=4$ samples are received on each input clock, $2n/4$ input clocks are needed for each memory write.  From this we find that
$$
                f_s \ge {{n/2}\over{w+1}}f_i = {{n^2} \over {4N+2}}f_i
$$
where $f_i$ is the input clock (INCLK) rate and $f_s$ is the memory clock (\sysclk) rate.  If \sysclk\ is faster than this minimum, the address generator automatically causes the extra cycles to be refresh-only cycles for the memory and idle cycles for the CMACs.

The CMAC clock is a gated version of \sysclk\ so that it is active only during memory read cycles.  Otherwise all CMACs are idle and consume no dynamic power.

After each SI, 8192 cycles of the output clock are needed to read out the 16b+16b complex results from the 4096 CMACs over the 16\,b DATAOUT bus.  This must be done before the next SI is complete, which means that
$$
                f_o \ge 8192{{w+1} \over {wT}}f_s
$$
where $f_o$ is the \outclk\ rate.

Both \sysclk\ and \outclk\ can be synthesized internally from INCLK using a PLL in the clock generator, but this is subject to some constraints.  There is only one PLL; it contains an internal delay-controlled oscillator in the range 3--6 GHz locked to $f_i R/M$, where $R$ and $M$ are integers determined by user-settable registers.  The oscillator drives two separately-programmable frequency dividers, one of which generates \sysclk.  Then \outclk\ is either generated by the other divider or supplied by the user at input pin USROUTCLK.  The two divider ratios and the selection of the \outclk\ source are configurable via the SPI interface.

\subsection{Address sequence}\label{sec:addrseq}

The memory word size of 1024\,b or 128 samples was selected so that there is one memory read cycle for each CMAC cycle.  However, one CMAC cycle requires 64 row samples and 64 column samples, so it would have been more straightforward to have 512\,b (64 sample) memory words and perform two reads per CMAC cycle, one for the row data and one for the column data, which are always at different locations in memory (Fig.~\ref{fig:subint}).  Then the maximum CMAC rate would be limited to half the maximum memory rate, which would limit performance because the CMACs can operate much faster.  To avoid this limitation, we use wider memory words, but then some data reordering is needed between the memory and the CMAC array.  That is the purpose of the buffer module (Fig.~\ref{fig:block2}).  

Each memory word contains simultaneous time samples from $n=64$ different signals along with the next time samples from the same 64 signals.  Each subsequent word contains two more time samples, so that $T/2$ words contain all $T$ samples for those signals.  Similar blocks of $T/2$ words hold $T$ samples for other groups of $n$ signals, with $w=2N/n$ groups altogether (Fig.~\ref{fig:subint}).  Reading from memory to buffer is done in sets of 4 successive clock cycles, obtaining time samples $t$ through $t+3$ for the row signals and the column signals:
\begin{unnumlist} 
\item {\it Memory to Buffer}
\item  \quad row data containing samples for $t$ and $t+1$
\item  \quad row data for $t+2$ and $t+3$
\item  \quad column data for $t$ and $t+1$
\item  \quad column data for $t+2$ and $t+3$.
\end{unnumlist}
On the next 4 clock cycles, the same data are transferred from the buffer to the CMAC array for processing, but in a different order:
\begin{unnumlist}
\item {\it Buffer to CMAC Array} 
\item \quad row $t$ and column $t$
\item \quad row $t+1$ and column $t+1$
\item \quad row $t+2$ and column $t+2$
\item \quad row $t+3$ and column $t+3$.
\end{unnumlist}
The buffer size is 8 memory words (double buffering) so that memory reading and CMAC processing can be overlapped with a latency of 4 clock cycles.

Processing uses sets of 4 words rather than 2 in order to ensure that no memory bank is addressed on two successive cycles.\footnote{This constraint arises because of internal pipelining in the memory.  A version of the DRAM without pipelining could have been used, avoiding the constraint, but it would then have been considerably slower, limiting chip performance under some circumstances.  Performance of the current design is always limited by I/O bandwidth, not by the speed of the DRAM or CMACs.}  The 16\,b memory address is structured as\hfill\break
\vskip -10pt
\centerline{address[15:0] = \{withinBank[10:0], bankNumber[4:0]\}.}
\smallskip
\noindent By always reading two adjacent addresses on successive cycles, we are assured that those cycles access different banks.

\begin{figure}[h]
%\vskip -0.10in
\begin{center}
\includegraphics[height=2.5 in, trim=0in 7.2in 1.7in .75in, clip]{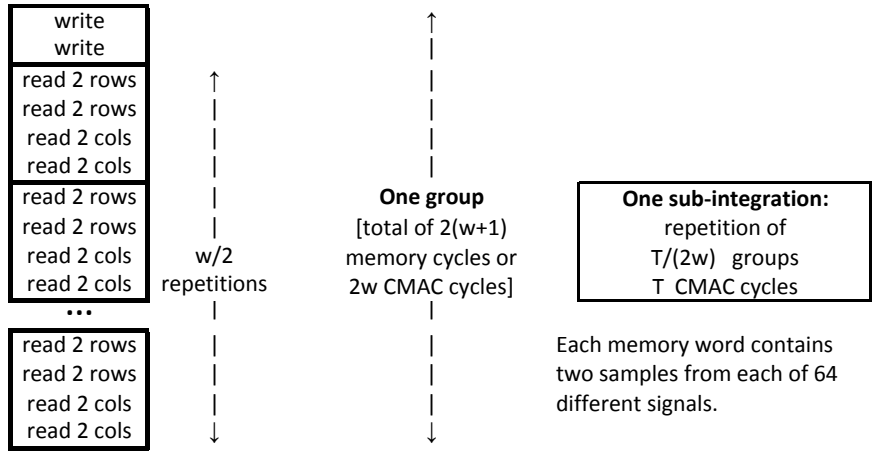}
\end{center}
\vskip -0.25in
\caption{Memory address sequence.  Two memory words at adjacent addresses are written or read in succession.  A group consists of 2 write operations and $2w$ read operations, achieving the required ratio of writing and reading.   A sub-integration is complete after processing $T/2w$ groups.}
\label{fig:addrseq}
\end{figure}

As explained earlier, a write cycle for new data must occur for every $w$ read cycles.  The shortest possible complete sequence then includes $w/2$ sets of four read cycles ($2w$ cycles) and two write cycles, as illustrated in Figure \ref{fig:addrseq}.  
The two write cycles use adjacent addresses, so the bank access constraint continues to be satisfied.  The sequence then includes $2(w+1)$ cycles, and $T/2w$ such sequences are needed to complete a sub-integration.

This structure introduces two constraints on user parameters.  First, $w$ must be even so that the sequence can include $w/2$ sets of read cycles.  This means that $N$ must be a multiple of $n=64$.  Second, $T$ must be a multiple of $2w$ because each sequence contains $4\times w/2 = 2w$ time samples.

The constraint that $N$ must be a multiple of 64 is mitigated for small numbers of antennas by a special ``memory bypass mode'' that supports $N=32$ (64 signals) by bypassing the memory entirely and sending input data directly to the CMAC array.  Two sets of 64 signals are processed in parallel, e.g. from different frequency channels.  Bypass mode is selected by a bit in a user-controlled register.   Thus, the chip can efficiently support $N=32$, 64, 128, 192, $\ldots$.  For other values of $N$, dummy signals (preferably always zero) must be added%
\footnote{It would be possible to avoid the requirement that $w$ be even and thus allow $N$ to be any multiple of 32 by lengthening the processing sequence.  If memory writing were done in sets of 4 addresses rather than 2, then each sequence would include $w$ read sets and $w$ could be any integer.  This would exacerbate the constraint on $T$, which would then need to be a multiple of $4w$.}.

On each cycle, the memory is given not only an address for reading or writing, but also a concurrent refresh bank number.  The latter must be different from the bank in the current, previous, and next read or write address, as well as different from the current, previous, and next refresh bank.  It is tricky to meet this requirement while ensuring that all banks get refreshed sufficiently often.  Our algorithm for this is not necessarily optimum, but it is reasonably efficient in that it requires a clock rate only about 13\% higher than an ideal one that distributes all cycles uniformly among the banks.

\subsection{Synchronization}

Once the appropriate memory addresses are written to the control registers, the chip processes each sub-integration automatically, delivering the results at the OUTDATA pins over 8192 cycles of CLKOUT.  Output signal OUTSYNC is asserted when the first word of each SI is valid.  The frequency of OUTCLK must be high enough to deliver all 8192 words before the next SI is completed.  OUTCLK can be faster, in which case more than 8192 cycles will occur before the next assertion of OUTSYNC, but only the first 8192 output words are valid.  The internal processing runs on \sysclk, whose frequency must be fast enough to keep up with the input data.  It can be faster, in which case idle cycles are automatically inserted by the address generator module.  In this way, the overall speed of operation is determined entirely by INCLK.  

%Input signal INTEGRATE must be asserted during the cycle of INCLK when the first input word of a full integration is delivered.  This forces the internal logic to begin of a new SI.  Another SI starts automatically every $T$ CMAC cycles thereafter.  An integration must consist of an integer number of SIs (namely $w^2/2$), so each time that INTEGRATE is subsequently asserted the chip should already be starting a new SI; if not, an error bit is set in one of the user-readable registers, indicating an inconsistency between the register-specified value of $T$ and the period of INTEGRATE or a \sysclk\ frequency that is too low.  In principle, INTEGRATE need only be asserted for the first integration; operation will then proceed correctly provided that the INDATA sequence begins a new block of $2NT$ samples every $2NT/4$ cycles of INCLK.

%Other than checking the timing of INTEGRATE as just described, 
The chip does not automatically keep track of full integrations; it operates at the SI level.  To establish the integration-level sequence (Figs.\ \ref{fig:subint}, \ref{fig:addrplot}), the user must supply the starting memory addresses of the row data, column data, and input (write) data for each SI, along with a bit that specifies whether this SI uses split mode.  This is done by writing to particular registers over the SPI link.  Values written during the current SI are used during the next one.  Writing may be begin immediately after OUTSYNC is asserted, and must be completed before the current SI is finished, as indicated by the next assertion of OUTSYNC.

Updating the addresses for each SI could have been further automated within the chip rather than putting that burden on the user, but the chosen design ensures that the chip has great flexibility.  It allows a wide range of $N$ to be supported, and it allows a different memory address sequence than that shown in Fig.\ \ref{fig:addrplot} to be used if desired.  A somewhat different sequence, discussed in section \ref{sec:largeN}, is useful when $N$ is sufficiently large; this permits the chip to support arbitrarily large $N$.

Internal synchronization signals are needed to account for the processing latency of each block in the signal flow.  These are shown as blue lines in Fig.\ \ref{fig:block2}.  The input module requires 32 CLKIN cycles to collect enough data to write one 1024-b word to the memory.  At the beginning of an integration, it delivers a delayed version of INTEGRATE to the address generator module when the first word is ready to be written to the memory.  The address generator then provides a signal marking the start of the first SI as well as subsequent SIs.  This signal is passed through the memory module, where it is delayed by the memory's read latency (but not otherwise used), and similarly through the buffer module, and then to the CMAC array, where it forces each CMAC to copy its accumulator to its readout register and to ready the accumulator for the next SI (Fig.\ \ref{fig:cmac2}).  The synchronization signal is then passed to the output module, where it is delayed by the output latency and delivered to the SYNCOUT pin at the same time as the first output word of the SI is delivered to DATAOUT.

%Due to the overlapping of writing new data to the memory with computing of the current integration, the SI that begins just after INTEGRATE is not part of the integration whose data are being received at DATAIN; it belongs to the previous one.  If we number the SIs from 1 to $w^2/2$, then the one then being computed is $w+1$.  During computation of this SI, the previous one, $w$, is being delivered to the output.  Thus the overall latency from the start of a new integration at the input to the start of delivering that integration's data to the output is $w^2/2 - w + 1$ SIs plus internal chip latencies equal to a few cycles of each clock.  The user must take these latencies into account to understand the order of the output data and to provide the proper memory addresses for each SI.  

\subsection{Inputs and Outputs}

The selection of parallel, synchronous bit streams for the main data input and output is a compromise driven by cost.  Simulations \cite{lrd2015c} in SPICE \cite{spice} have shown that at least 500 Mb/s can be transferred between each pin and a modern FPGA via a wire-bond package and 76 mm-long printed circuit trace.  Somewhat higher rates may actually be achieved, but 500 Mb/s has been used for the performance calculations of this paper.  The number of pins is limited by package cost and size as well as timing skew.  We selected 32 input pins and 16 output pins, providing 16 Gb/s input bandwidth and 8 Gb/s output bandwidth.  These bandwidths provide good performance for all $N \ge 32$, as shown in section \ref{sec:PvsN} and Fig.\ \ref{fig:performance}.  To achieve higher rates would require multi-Gb/s serial transmitters and receivers.  Such devices are technically feasible, but including them in the present design is cost prohibitive.  (An {\it ab initio} design would be difficult and risky for our small team, and existing designs command high license fees.)  Larger input bandwidth would permit a higher processing rate at small $N$, and larger output bandwidth would do so at large $N$.  This would reduce the chip count in a large system, but it would have little effect on the system power consumption.

The design includes 39 input pins and 19 output pins (Fig.\ \ref{fig:block2}), not including power, ground, and test connections.  All are implemented with the same bi-directional I/O cell from IBM.  It provides a low-voltage CMOS interface at a nominal voltage swing of 0.9 V, which is the same as the nominal supply voltage of the internal logic.  Each cell is configured internally as either input or output, since none of our pins is bi-directional.  For outputs, it is configured for maximum slew rate and 50 ohm source resistance.  Additional discussion of these cells is given in Section \ref{sec:iopower}.

\subsection{Control Interface}

The chip is controlled by a set of 12 registers that can be written or read over a 4-wire SPI slave interface \cite{spi}.  It is intended to be driven by a corresponding SPI master interface in an external controller, such as an FPGA.  Multiple correlator chips on the same board can be connected to the same controller; one of the wires (SSEL) selects the current correlator chip.  

Each register contains 20 bits.  Our protocol uses 25 SPI clock cycles per transfer (4 address bits, one write enable bit, and 20 data bits).  On each transfer, all bits of the addressed register are read and returned over the SPI MISO (master in, slave out) wire.  Data sent on the MOSI (master out, slave in) wire are then written to the register if the write enable bit is set.  A portion of one register has a set of read-only status bits for reporting error conditions.  Once set by the internal logic, the status bits remain set until that register is read, then they are automatically cleared.  Register reading and writing are driven entirely by the SPI clock (wire SCLK), generated by the master interface, independent of any other clocks.  This ensures that they can be set even if the internal PLL is not running or is unstable because it was previously misprogrammed or because INCLK is not yet being supplied.  

At power-up or after assertion of active-low RESETN, the registers are loaded with default values which allow operation to proceed.  Unless the defaults happen to correspond to the present application, the output data will not be valid until the registers are written with the appropriate values.  These include programming the PLL to generate \sysclk\ at a frequency high enough to keep up with the input data, and specifying the values of $w$ and $T$, which determine the addressing sequence (Fig. \ref{fig:addrseq}) and the SI length.  In most applications, these settings need only be written during initialization.

Three registers must be written during each sub-integration.  These give the starting addresses for column data reading, row data reading, and new data writing during the next SI, and whether the next SI uses split mode.  This requires 75 cycles of SCLK, and must be completed during the time of one SI, which is $T$ CMAC clocks, so there is a minimum value of $T$ that varies inversely with the SPI clock rate and the CMAC rate.  The maximum CMAC rate is 312.5 MHz at $N=640$ (see Section \ref{sec:PvsN}).  At an SCLK clock rate of 15 MHz, this gives $T>1718$.  The memory size is sufficient to support $T=5480$ at $N=640$.  The CMAC rate and maximum $T$ are such that an SCLK rate $<15$ MHz is sufficient at all $N$, and the chip will support an SCLK rate of at least 25 MHz.

\section{Synthesis and Simulation}

\subsection{Methodology}

The register-transfer-level (RTL) design was created in Verilog.  Three vendor-specific hard-macro cells were instantiated:
\smallskip
\begin{list}{--}{\topsep 0pt \leftmargin 2pc \itemindent -2\labelsep}
\item DRAM, $512\times 65536$, a black-box macro from IBM.  Two of these are used in parallel inside the memory module to create a $1024\times 65536$ memory.
\item PLL, also a black-box macro from IBM, used in the clock generator module.
\item Bi-directional LVCMOS input/output cell, from IBM.
\end{list}
\smallskip
The design was synthesized using standard cell libraries from ARM\footnote{ARM Ltd., product SMC9MC, HVt, SVt, and UVt versions \cite{arm}.} by means of Synopsys Design Compiler\footnote{DC Ultra, Version K-2015.6 \cite{synopsys}.}, with timing constraints based on $f_i=500$~MHz, $f_s=360$~MHz, and $f_o=500$~MHz.  Except for its use of the above cells from IBM, the RTL design is portable and could be synthesized for fabrication in a different process by using an appropriate standard-cell library.

Library files giving timing and power data on all standard cells and on the above special cells were available for various process-voltage-temperature corners.  During synthesis, we used a ``slow'' corner at drain-source voltage 0.8\,V ({\it cf.} 0.9\,V nominal) and temperature --40\,C.  The resulting synthesized netlist and the calculated delays of all cells and nets were then used by ModelSim\footnote{ModelSim SE-64 version 10.6c for Linux \cite{modelsim}.} to simulate operation of the chip for a specific scenario ($N=128$ antennas and integration length $T=1032$) and a specific set of simulated input signals.  The simulation test bench collected the chip's output data and these were later compared against independently-calculated correct results.  The simulator also calculated the switching activity on each net.  The switching activity was read back into Design Compiler and its Power Compiler component was used to calculate the power dissipated by each cell.  For the latter calculation, we used a more practical corner (0.9~V and +50~C) in order to obtain realistic switching and leakage power estimates.  

Three versions of the standard cell libraries were available, each containing cells of the same functionality but at different CMOS threshold voltages.  The compiler was told that cells from the highest-threshold library are preferred, since this minimizes leakage power, but it was allowed to select up to 10\% of the cells from the lower-threshold libraries when optimizing timing.  The final result used 6.17\% from the low-threshold libraries.

Design Compiler was used in its Topographic Mode, where it makes use of data about the process technology (metal layer stackup) and physical data about all the cells (dimensions and port locations) to perform a rough placement of the cells.  This allows it to make accurate estimates of the delay and capacitive load of each wiring net.  Although a complete layout has not yet been done, we believe that this provides reliable power estimates.

The results of this process are summarized in Table \ref{tbl:pcresults}.
\begin{wstable}[h]
\caption{Preliminary Power and Cell Area, $N=128$ and $T=1032$}
\begin{tabular}{@{}lrrrrrrrr@{}} 
\toprule
\hfill Module\hfill\hfill & \multicolumn{4}{c}{\hrulefill\ Power, mW\hrulefill}            & Area, mm$^2$ \\
                   & Switching    & Internal & Static & \hfill Total\hfill\hfill & \\
\colrule
CMAC Array&129.199&179.727&36.500&345.386&2.728296 \\
Memory&2.970&{\it 123.994}&10.000&136.983&6.392190 \\ 
Input cells (39)&{\it 3.702}&{\it 0.331}&0.286&3.257&0.200353 \\
Output cells (19)&{\it 33.391}&{\it 6.248}&0.139&39.774&0.131266 \\
Clock Generator&0.875&0.013&54.000&54.880&0.078929 \\
Control&0.000&0.005&0.006&0.012&0.001333 \\
Output&12.943&3.766&5.490&22.196&0.102880 \\
Buffer (DRAM to CMACs)&2.024&7.860&0.572&10.457&0.034140 \\
Address Generator&0.021&0.116&0.016&0.153&0.000704 \\
Input&2.051&6.650&0.286&8.988&0.012865 \\
Misc. (top module)&3.981&0.000&0.695&4.530&0.07335 \\
\colrule
Total of all modules & 191.156&327.619&107.989&626.617&9.756312 \\
\botrule
\end{tabular}
\begin{tablenotes}
\item Clock buffer trees not included, see Section \ref{sec:clockTrees}
\item Values in italics are subject to corrections described in Section \ref{sec:corrections}; final values are given in Table \ref{tbl:corrected}.
\end{tablenotes}
\label{tbl:pcresults}
\end{wstable}
As expected, a majority of the power is used by the CMAC array (55\%) and the memory (22\%).   Static power, mostly due to cell leakage currents, is low.  Nearly all of the power of the clock generator is static, but in that case it is not leakage but rather is due to the PLL's internal microwave oscillator, which runs even without input switching activity.  More than 65\% of the cell area is occupied by the memory; the design would be impractical if lower-density memory ({\it e.g.}, static RAM) were used.

We have found that several small corrections to these results are necessary.  First, the clock networks were treated as ideal during synthesis, so buffer trees for them were not synthesized.  Therefore, we carried out a rough, manual synthesis of the clock trees for the purpose of estimating their power consumption and area.  Second, the simulator did not provide enough switching activity detail to determine accurately the internal power of the memory.  This had only a small effect in the simulated scenario, but it is significant when the results are extrapolated to other scenarios, as we do in section \ref{sec:PvsN}.  Third, the I/O cell power estimates from Power Compiler are inaccurate for several reasons, so we re-computed those manually.  This produced only a small net change.  Details of these corrections are described sections \ref{sec:clockTrees} and \ref{sec:corrections} and the final results are given in section \ref{sec:results}.

\subsection{Post-Synthesis Simulation}\label{sec:sim}

The compiler generated a synthesized netlist and a Standard Delay Format \cite{sdf} file containing the signal delays through all cells and nets.  These were used with a Verilog test bench and simulated input data to compute a detailed simulation under realistic conditions.  The case simulated was $N=128$ antennas (256 input signals) and integration length $T=1032$.  This integration length is the minimum that does not require an output clock faster than 500 MHz.  Actual integration lengths can be much longer and $N$ can be much larger, but simulations in such cases would have impractical run times.

The simulated input data consisted of quantized Gaussian noise, independent among all signals and among all samples of each signal, with standard deviation $\sigma = 3.0$ for the real and imaginary parts.  An additional time series with independent samples and $\sigma = 3.0$ was added to two of the signals, creating a cross-correlation coefficient of 0.5 for that pair but zero for all other pairs.

The test bench generated the INCLK signal at 500 MHz and used a simulated SPI master controller to load parameters into the chip's registers, just as will be done in the actual device.  These parameters commanded the PLL to generate the internal clock (\sysclk) at $f_s=227.273$ MHz and the output clock at $f_o=500.0$ MHz, and they set $N$ and $T$ to the selected values.   The test bench then loaded the pre-computed input samples in the appropriate sequence.

Under these conditions, the average CMAC rate is 62.5 MHz, and $f_s=78.125$ MHz would be sufficient to keep up with the input data.  The higher $f_s$ rate was used to ensure that the internal DRAM has sufficient refresh cycles.  (No extra cycles for refresh are needed when  $N \ge 448$.)

For $N = 128$, 8 sub-integrations (SIs) are needed for each integration.  The simulated input data included 24 SIs covering 3 integrations, with each integration repeating the same data values.  The output data are invalid until sufficient input data have been written to the internal memory.  The first integration is read out as the 6th through 13th SIs, the second as the 14th through 21st, and the third is only partly read out (first 3 SIs) because the simulation ends when the input data are exhausted.  The test bench collects all the output data and saves it to a file, recording 24 SIs of which the last 19 are valid.

%The post-synthesis simulation (netlist with delays) did not always give perfect results.  There were no output errors with the constant-value data, but with the Gaussian noise data about 67\% of the complex output values were wrong.  The same 33\% were correct in each SI.  Detailed investigation showed that the errors resulted from incorrect values being read from the DRAM.  An additional simulation was run using the synthesized netlist but with all delays assumed to be zero.  This was free of errors, showing that the errors are very likely to be caused by timing errors, even though the compiler found no setup timing violations during synthesis.  The RTL code was also simulated with the same input data, 
%
%The errors are likely due to the fact that the IBM-supplied library for the DRAM contains no timing information except for some test pins that are not used in normal operation.  No setup or hold constraint is given for the input pins (control and data), and no clock-to-output delay is given for the data output pins.  Typical values are given on the printed data sheet, but we do not have values at multiple PVT corners as we do for other cells.  This makes it impossible for the compiler to ensure proper timing at the inputs and outputs of the DRAM.  Similar errors probably occur for the constant-value data, but we did not check in detail.   For that case, errors can be masked by the fact that the same values are used for 1032 successive samples and that the 4 LSBs of the CMAC results are rounded off at the chip output. 

The simulation ends after at 419.793\,$\mu$s of simulation time.  From simulation time 110\,$\mu$s to 210\,$\mu$s, the simulator records the switching activity on all nets in a Switching Activity Interchange Format (SAIF) file\footnote{See http://www.synopsys.com/Community/Interoperability/Pages/TapinSAIF.aspx .}.  This covers the time when SIs 6 through 11 were being computed and read out; it avoids all initialization as well as all invalid SIs and thus represents steady-state operation of the chip.

\subsection{Clock Trees}\label{sec:clockTrees}

There are three large clock networks that were treated as ideal during synthesis, as listed in Table \ref{tbl:clocks}.
\begin{wstable}[h]
\caption{Networks Treated As Ideal}
\begin{tabular}{@{}cccc@{}} 
\toprule
Net & Fanout & Load, fF \\ 
\colrule
INCLK\hfill         &  \hphantom{00}3,168   & \hphantom{00}1,626 \\
sysclk\hfill          &  \hphantom{00}8,352  & \hphantom{00}4,463 \\
CLK\_CMAC\hfill &                     311,488  &                     167,800 \\
\botrule
\end{tabular}
\label{tbl:clocks}
\end{wstable}
In view of the large fanouts, each will need a tree of buffer cells, but these were not synthesized by the compiler.  It is best to create the final clock trees during layout, when their timing can be more precisely optimized.  

To estimate the power consumption and area of the three clock trees, we performed a rough, manual synthesis of each.  The resulting designs are far from optimum, so they represent an upper limit on the required power.  We selected (somewhat arbitrarily) two standard cells, an inverting and a non-inverting buffer, from which to build the clock trees.  We adopted the constraint that the delay though any path of the tree should be no more than 20\% of the minimum clock period for that net.  Details of the calculations are given in Appendix A and the results are summarized in Table \ref{tbl:trees}.  The switching power for driving the input pins of all destination cells of each net was included in the Power Compiler results, so that power is subtracted from the total to determine the net power added by each buffer tree.
\begin{wstable}[h]
\caption{Clock Buffer Trees}
\begin{tabular}{@{}lrrrrrrrr@{}} 
\toprule
\hfill\hfill Net:&CLK\_CMAC &\hfill sysclk\hfill\hfill & INCLK &\hfill Totals\hfill\hfill \\
\colrule
Max.~frequency, MHz\tnote{a}         & 360            & 360      & 500  & \\
Frequency, MHz\tnote{b}                 &62.5            & 227.273 & 500 & \\
Total delay, ps                    &494.6& 521.2& 332.2& \\
Total power, $\mu$W          & 44236.23   & 4263.47 & 3674.67 & 52174.36 \\
Load pin switching, $\mu$W& 10599.91  & 1039.17 &  884.53 & \\
Net power added, $\mu$W       & 33636.32   & 3224.30 & 2790.14 & 39650.76 \\
Number of tree levels              &      3          &     2      &    2\\
Total number of buffer cells      & 2762         &   49      & 27   & 2368 \\
Total cell area, $\mu\rm m^2$ &  3556.1& 63.1& 34.7& 3653.9\\

\botrule
\end{tabular}
\begin{tablenotes}
\item[a] Maximum switching frequency of net over all $N$, for delay constraint.
\item[b] Effective switching frequency in the simulated scenario, for power calculation.
\end{tablenotes}
\label{tbl:trees}
\end{wstable}

\subsection{Other Corrections}\label{sec:corrections}

\subsubsection{I/O cells}\label{sec:iopower}

For the I/O cells, we have manually estimated the power consumption using data from the cell library files.  These results are shown in Table \ref{tbl:iopwr} for the simulated scenario.
\begin{wstable}[h]
\caption{I/O Cell Power}
\begin{tabular}{@{}lcrrrrr@{}} 
\toprule
\quad Pin name& Count &Toggle Rate&\multicolumn{4}{c}{\hrulefill\ Power, mW \hrulefill} \\
                   &          &\hfill MHz\hfill\hfill& Switching& Internal &\hfill Static\hfill\hfill&\hfill Total\hfill\hfill \\
\colrule
{\it Input Pins} \\
INCLK&1&1000.00\hphantom{00}&0.72576&0.10150&0.00706&0.83432 \\
INDATA&32&47.78\hphantom{00}&0.00642&0.80480&0.22580&1.03702 \\
INTEGRATE&1&0.00\hphantom{00}&0.00000&0.00000&0.00706&0.00706 \\
SPI\_MOSI&1&0.00\hphantom{00}&0.00000&0.00000&0.00706&0.00706 \\
SPI\_SCLK&1&0.00\hphantom{00}&0.00000&0.00000&0.00706&0.00706 \\
SPI\_SSEL&1&0.00\hphantom{00}&0.00000&0.00000&0.00706&0.00706 \\
RESETN&1&0.00\hphantom{00}&0.00000&0.00000&0.00706&0.00706 \\
USROUTCLK&1&0.00\hphantom{00}&0.00000&0.00000&0.00706&0.00706 \\
{\it Output Pins} \\
OUTCLK&1&1000.00\hphantom{00}&6.73313&2.28490&0.00706&9.02508 \\
OUTDATA&16&250.00\hphantom{00}&26.93250&9.13960&0.11290&36.18500 \\
OUTSYNC&1&0.12\hphantom{00}&0.00082&0.00028&0.00706&0.00816 \\
SPI\_MISO&1&0.00\hphantom{00}&0.00000&0.00000&0.00706&0.00706 \\
\colrule
TOTALS&58&&34.39863&12.33108&0.40925&47.13896 \\
Input pins& 39 &&0.73218&0.90630&0.27519&1.91367 \\
Output pins& 19 &&33.66645&11.42478&0.13407&45.22529 \\
\botrule
\end{tabular}
\label{tbl:iopwr}
\end{wstable}
The leakage power is essentially the same for all cells and is about 1\% of the total.  The dynamic power is dominated by the clock and data pins, since the remaining 8 pins have negligible switching activity.  Most of the power is used by output cells for driving the chip's outputs.  We assumed a load of 15.3 pF for each output pin, derived from our model of a flip-chip package, PC board track, and FPGA input pin.  Pin USROUTCLK was not used in the simulated scenario; if it were used at 500\,MHz, it would add about 1\,mW.

The total for all cells is in good agreement between the two estimates (43 mW in Table \ref{tbl:pcresults} and 47\,mW in Table \ref{tbl:iopwr}), but there are significant discrepancies in the details.  Power Compiler's estimate of the internal power of the cells is about half of our estimate, but the switching power is higher.  This is the result of several issues.  First, Power Compiler treated each top-level port as bi-directional, so switching power is included for its output function, even if it is really just an input.  Second, the internal power of the cell is a strong function of the transition time of its input signal (in both directions).  For inputs, we assumed realistic 500\,ps transitions, whereas we told Design Compiler and Power Compiler that the input sources have infinite drive strength and zero transition time.  For outputs, we assumed 50\,ps transitions at the cell inputs, which may have been too pessimistic.  
%For unknown reasons, Power Compiler gave ``N/A" as the internal power of the INCLK I/O cell.

We resolve these discrepancies by adopting our estimates of the I/O cell power.  This provides accurate values for switching and leakage power (which are simple to calculate and comprise 74\% of the total) and conservative estimates of the internal power due to use of pessimistic transition time estimates.

\subsubsection{Memory cycle distribution}\label{sec:mempwr}

There is one more difficulty; it affects only the internal power of the memory module, which is 19.8\% of the total in Table \ref{tbl:pcresults}.  The SAIF file gives, for each net, the total time at logic 0, the total time at logic 1, and the number of transitions in the 100\,$\mu$s analysis period.  For some cells, this is not sufficient for calculating their internal power because the energy dissipated due to a transition on one pin (such as a clock) depends on the states of other pins.  For almost all cells this is a negligible effect, but it is important for the DRAM, where the energy is different in read, write, read+refresh, write+refresh, refresh only, and idle cycles.  ModelSim is not capable of reporting state-dependent switching activity.  In that case, Power Compiler assumes that the switching activity at the various pins is statistically independent with the probabilities given by the logic 0 and 1 times in the SAIF file.  For the particular scenario that we have simulated, it turns out that this is a good approximation to the actual activity, but we now calculate a correction.  This will be more important when the results are extrapolated to other scenarios (section \ref{sec:PvsN}).

The DRAM has three active-low control pins, READN, WRTN, REFN, which specify, respectively, a read cycle, a write cycle and a refresh cycle.  Thus in principle there are 8 possible cycle types, but two are excluded because reading and writing on the same cycle is not possible.  Of the remaining ones, our design uses only three:\hfill\break
\vbox{
\smallskip
\begin{tabular}{@{}ll@{}}
  !READN \& WRTN \& !REFN &  read with concurrent refresh,\\
  READN \& !WRTN \& !REFN &  write with concurrent refresh,\\
  READN \& WRTN \& !REFN  &  refresh only.\\
\end{tabular}\smallskip
}\hfill\break
For the simulated scenario, the fraction of clock cycles in each state and the implied average frequencies are:\hfill\break
\vbox{
\smallskip
\begin{tabular}{@{}llr@{}}
read+refresh  &  0.275\hphantom{00}    &62.500 MHz,\\
write+refresh & 0.06875                        &15.625 MHz,\\
refresh onlly   & 0.65625                       &149.148 MHz,\\
\qquad Total (\sysclk\ rate)  &  1.00000  &227.273 MHz.\\
\end{tabular}\smallskip
}\hfill\break
As expected, the SAIF file implies probabilities for each control pin of\hfill\break
\vbox{
\smallskip
\begin{tabular}{@{}ll@{}}
  !READN  & $0.27500 = p_r$,\\
  !WRTN  & $0.06875 = p_w$,\\
  !REFN  &  $1.00000$.\\
\end{tabular}\smallskip
}\hfill\break
Although the pin states are not independent, Power Compiler assumes, in the absence of any other information, that they are.  There are then 4 possible states with probabilities\hfill\break
\vbox{
\smallskip
\begin{tabular}{@{}lcrl@{}}
!READN \& !WRTN &  (invalid)   &          $p_r p_w =  .018906$,\\
!READN \& WRTN  & read         &     $p_r (1-p_w) =  .256093$,\\
READN \& !WRTN  &  write       &      $(1-p_r) p_w = .048744$,\\
READN \& WRTN  &   refresh    &  $(1-p_r)(1-p_w) = .675156$.\\
\end{tabular}\smallskip
}\hfill\break
This implies rates of\hfill\break
\vbox{
\smallskip
\begin{tabular}{@{}lr@{}}
!READN \& !WRTN   &     4.297 MHz,\\
!READN \& WRTN    &   58.208 MHz,\\
READN \& !WRTN    &   11.328 MHz,\\
READN \& WRTN    &  153.445 MHz.\\
\end{tabular}\smallskip
}\hfill\break
If we multiply theses rates by the corresponding energy per clock transition for the appropriate corner, we get a total of  60.953 mW of internal power per DRAM cell.  Additional internal power is dissipated per output pin transition; since the data are random, those pins should have transitions on 50\% of the read cycles; this is in fact what the SAIF file shows.   Applying this to the energy per transition for all 512 output pins gives an additional power of 0.452 mW.  There are two DRAM cells in the memory module, giving 122.810 mW altogether, compared to 123.994 mW in Table \ref{tbl:pcresults}.  The difference can be attributed to Verilog wrapper logic provided by IBM.

In view of the good agreement in this analysis, we can proceed to correct the value in Table \ref{tbl:pcresults} by using the actual rates for each state.  This gives 130.604 mW for the memory module (including 1.184 mW for the wrappers), 5.3\% higher than in Table \ref{tbl:pcresults}.  More importantly, it provides a breakdown of power by the type of memory cycle, enabling us to extrapolate to scenarios where the state probabilities are different.

\subsection{Results}\label{sec:results}

\subsubsection{Power}

Using the clock tree power and area from Table \ref{tbl:trees}, the I/O cell power from Table \ref{tbl:iopwr}, and the corrected memory internal power from Section \ref{sec:mempwr} along with all other results from Table \ref{tbl:pcresults} gives the final totals in Table \ref{tbl:corrected}.
\begin{wstable}[h]
\caption{Final Power and Cell Area Estimates\tnote{a}}
\begin{tabular}{@{}lrrrrr@{}} 
\toprule
\quad Module& \multicolumn{4}{c}{\hrulefill\ Power, mW \hrulefill}  & Area, mm$^2$ \\
 & Switching & Internal & Static & Total & \\
\colrule
\it From Table \ref{tbl:pcresults}&  &  &  &  & \\ 
CMAC Array & 129.199 & 179.727 & 36.500 & 345.426 & 2.728296\\ 
Clock Generator & 0.875 & 0.013 & 54.000 & 54.888 & 0.078929\\ 
Input interface & 2.051 & 6.650 & 0.286 & 8.987 & 0.012865\\ 
Output interface & 12.943 & 3.766 & 5.490 & 22.199 & 0.102880\\ 
Control & 0.000 & 0.005 & 0.006 & 0.011 & 0.001333\\ 
Address Generator & 0.021 & 0.116 & 0.016 & 0.153 & 0.000704\\ 
Buffer (DRAM to CMACs) & 2.024 & 7.860 & 0.572 & 10.456 & 0.034140\\ 
\smallskip
Misc. (top module) & 3.981 & 0.000 & 0.695 & 4.676 & 0.073356\\ 
\it New or revised&  &  &  &  & \\ 
Clock buffers (Table \ref{tbl:trees}) & 38.639 & 0.985 & 0.027 & 39.651 & 0.003654\\ 
Memory\tnote{b}& 2.970 & 130.604 & 10.000 & 143.574 & 6.392190\\ 
Input cells (Table \ref{tbl:iopwr}) & 0.732 & 0.906 & 0.275 & 1.914 & 0.200353\\ 
Output cells (Table \ref{tbl:iopwr}) & 33.666 & 11.425 & 0.134 & 45.225 & 0.131266\\ 
\vspace{-4pt}
\colrule
Totals & 227.102 & 342.057 & 108.001 & 677.160 & 9.759966\\ 
\botrule
\end{tabular}
\begin{tablenotes}
\item[a] Power is for the simulated scenario:\hfill\break
\centerline{$N=128$, $T=1032$, $f_i = f_o = 500\,$MHz, $f_s=227.217\,$MHz.}
\item[b] For memory, only the internal power estimate is revised.
\end{tablenotes}
\label{tbl:corrected}
\end{wstable}

\subsubsection{Chip size}

Table \ref{tbl:corrected} gives the total area of all cells as 9.760 mm$^2$.   During layout, it will be necessary to add area between the cells to accommodate the wiring.  The compiler estimated that the total length of wires will be 40,993.6 mm.  The process provides 13 metal layers that can be used for wiring, but the uppermost layer provides the I/O pads and the next two layers are likely to be dedicated to core logic power and ground, leaving 10 wire routing layers.  These vary in minimum pitch from 0.1 to 0.8~$\mu$m, with an average of 0.19~$\mu$m.  Using the latter number with the estimated length gives an area of 7.789 mm$^2$ occupied by wires.  If this is spread uniformly among the 10 layers, it uses 0.7789 mm$^2$ of chip area.  Uniform use of the layers is unrealistic, and congestion in some places will no doubt require that the wiring be further spread out, so we double the latter value and estimate that 1.558 mm$^2$ of chip area will be needed for wiring.  Adding this to the cell area gives a total chip area of 11.318 mm$^2$.  This could be realized as a 3.4 mm square.

We next consider whether the chip might need to be still larger to accommodate I/O pads.  We are planning to use flip-chip packaging \cite{flipchip} with C4 bumps at 200 $\mu$m pitch, which is one of several options available from the foundry.  A dense array of 289 bumps could then be placed on the 3.4 mm square chip.  The design uses 58 pins for operating signals, and additional pins are needed for test signals as well as for power and ground, but the total is unlikely to exceed 100.  We conclude that the pins need not be densely packed and that no additional area needs to be included.

\section{Performance as a Function of $N$}\label{sec:PvsN}

Simulation of the $N = 128$, $T = 1032$ scenario for the synthesized netlist (Section \ref{sec:sim}) takes 10 to 15 hours on a multi-core Linux server with 32 GB of memory.  Under some circumstances, it can take much longer.  With larger numbers of antennas, it is necessary to increase the integration length $T$ to avoid excessive output clock rate, and the number of sub-integrations in a full integration is proportional to $N^2$, increasing the run time proportionally. It is therefore not practical to simulate large-$N$ or large-$T$ scenarios.\footnote{On the other hand, simulation of the RTL code with zero delay is much faster, so we were able to carry out a simulation that verifies functional correctness at $N=1024$.}

It is important to understand the performance at larger $N$ and $T$ because the chip is likely to be useful in those applications and because it then performs more efficiently.  
%For example, at $N \ge 448$ it is no longer necessary to have refresh-only cycles.  
Fortunately, the available simulation results can be accurately extrapolated to the larger cases.  The power estimates are broken down by major module and within each module by static and dynamic power.  The modules operate at different clock frequencies, but each has only one clock and we know how the clock rates vary with $N$ and $T$.  The dynamic power is proportional to clock rate and the static (leakage) power is independent of clock rate.  The memory module is somewhat more complicated; it runs on the internal clock (\sysclk), but it uses different amounts of energy in each cycle depending on whether it is a read cycle, a write cycle or a refresh-only cycle.  These energies are known from the cell library files, so each type of cycle can be extrapolated separately.

Table \ref{tbl:pwrVsN} shows such extrapolations for several values of $N$ from 256 to 2048.  
\begin{wstable}[h]
\caption{Power Use for Larger Numbers of Antennas}
\begin{tabular}{@{}lcrrrrrr@{}} 
\toprule
\hfill\hfill Number of antennas, $N$:&   &128&256&512&640&1024&2048 \\
\colrule
Maximum integration length & $T$ & 32768 & 14976 & 6944 & 5480 & 3328 & 1536\\
CMAC rate, Hz & $f_c$ & 6.250E+07 & 1.250E+08 & 2.500E+08 & 3.125E+08 & 2.031E+08 & 9.375E+07\\
Input clock frequeny, Hz & $f_i$ & 5.000E+08 & 5.000E+08 & 5.000E+08 & 5.000E+08 & 2.031E+08 & 4.688E+07\\
System clock frequency, Hz & $f_s$ & 2.270E+08 & 2.270E+08 & 2.656E+08 & 3.281E+08 & 2.270E+08 & 2.270E+08\\
Output clock frequency, Hz & $f_o$ & 5.000E+08 & 6.838E+07 & 2.949E+08 & 4.672E+08 & 5.000E+08 & 5.000E+08\\
Memory write rate, Hz & $f_w$ & 1.563E+07 & 1.563E+07 & 1.563E+07 & 1.563E+07 & 6.348E+06 & 1.465E+06\\
Idle (refresh only) rate, Hz & $f_r$ & 1.489E+08 & 8.638E+07 & 0.000E+00 & 0.000E+00 & 1.753E+07 & 1.318E+08\\
Bandwidth, Hz & $b$ & 7.813E+06 & 3.906E+06 & 1.953E+06 & 1.563E+06 & 3.967E+05 & 4.578E+04\\
\colrule
\vspace{3 pt}
{\it Power Estimates, mW} &          &{\it Simulated} & \multicolumn{5}{c}{\hrulefill\ {\it Extrapolated} \hrulefill} \\
CMACs switching + internal & $f_c$ & 308.926 & 617.852 & 1235.704 & 1544.630 & 1004.010 & 463.389\\
Memory switching & $f_c$ & 2.970 & 5.940 & 11.880 & 14.850 & 9.653 & 4.455\\
Memory internal - read & $f_c$ & 59.968 & 119.936 & 239.872 & 299.840 & 194.896 & 89.952\\
Memory internal - write & $f_w$ & 22.224 & 22.224 & 22.224 & 22.224 & 9.029 & 2.084\\
Memory internal - idle & $f_r$ & 47.230 & 27.402 & 0.000 & 0.000 & 5.560 & 41.808\\
Memory wrapper internal & $f_s$ & 1.182 & 1.182 & 1.383 & 1.709 & 1.182 & 1.182\\
Address  generator, sw+int & $f_s$ & 0.137 & 0.137 & 0.160 & 0.198 & 0.137 & 0.137\\
DRAM to CMAC, sw+int & $f_s$ & 9.884 & 9.884 & 11.566 & 14.287 & 9.884 & 9.884\\
Input interface, sw+int & $f_i$ & 8.701 & 8.701 & 8.701 & 8.701 & 3.535 & 0.816\\
Clock generator, sw+int & $f_i$ & 0.888 & 0.888 & 0.888 & 0.888 & 0.361 & 0.083\\
Control, sw+int & $f_i$ & 0.006 & 0.006 & 0.006 & 0.006 & 0.002 & 0.001\\
Output interface, sw+int & $f_o$ & 16.709 & 2.285 & 9.856 & 15.611 & 16.709 & 16.709\\
Misc. (top module), sw+int & $f_s$ & 3.981 & 3.981 & 4.658 & 5.754 & 3.981 & 3.981\\
Static, all above modules & 1 & 107.565 & 107.565 & 107.565 & 107.565 & 107.565 & 107.565\\
Input cells (Table \ref{tbl:iopwr})& $f_i$ & 1.638 & 1.638 & 1.638 & 1.638 & 0.666 & 0.154\\
Output cells (Table \ref{tbl:iopwr}) & $f_o$ & 45.091 & 6.166 & 26.598 & 42.129 & 45.091 & 45.091\\
I/O cell leakage & 1 & 0.409 & 0.409 & 0.409 & 0.409 & 0.409 & 0.409\\
CLK\_CMAC buffers (Table \ref{tbl:trees}) & $f_c$ & 33.610 & 67.221 & 134.442 & 168.052 & 109.234 & 50.416\\
sysclk buffers (Table \ref{tbl:trees}) & $f_s$ & 3.224 & 3.224 & 3.772 & 4.660 & 3.224 & 3.224\\
INCLK buffers (Table \ref{tbl:trees}) & $f_i$ & 2.790 & 2.790 & 2.790 & 2.790 & 1.133 & 0.262\\
Clock buffers leakage & 1 & 0.027 & 0.027 & 0.027 & 0.027 & 0.027 & 0.027 \\
\vspace{-4pt}
\colrule
Total, mW &  & 677.160 & 1009.385 & 1824.113 & 2255.936 & 1526.254 & 841.528 \\
\colrule
Energy per CMAC op, pJ & & 2.65 & 1.97 & 1.78 & 1.76 & 1.83 & 2.19 \\
\botrule
\end{tabular}
\label{tbl:pwrVsN}
\end{wstable}
The column for $N = 128$ uses results from Power Compiler (Table \ref{tbl:pcresults}) along with our estimates of the I/O cell power (Table \ref{tbl:iopwr}) and buffer tree power (Table \ref{tbl:trees}).  For the internal power of the memory module, the corrected results derived in section \ref{sec:mempwr} are used, separately for each cycle type.  At the top of the table, the average clock frequencies are given for each case.  For the memory, the read rate is the same as the CMAC rate ($f_c$), and the write rate is $f_w=f_c/w$, where $w=2N/64$.  The idle (or refresh-only) rate is $f_s - f_c - f_w$.  For each major module, the relevant clock is listed.  The dynamic power includes internal and switching power, both of which are proportional to the module's clock frequency.  The total static (leakage) power of all modules is given separately and is independent of $N$.  For the memory, internal and switching power are handled separately because switching power (to drive the output pins) is used only on read cycles.

The power FoM, energy per CMAC operation, is given at the bottom of Table \ref{tbl:pwrVsN}.  It is minimized at $N=640$, but it remains low over the entire range.  Except at small $N$, the majority of the power is dissipated in the CMACs; this is desirable, since all else can be considered overhead.  The CMACs themselves use 1.206 pJ per operation and 36.5 mW of static power in all cases (1.23 to 1.35 pJ per operation total).  

The results plotted in Fig.\ \ref{fig:performance} use the same method to calculate performance for additional values of $N$, including all multiples of 64 up to 4096, and also including $N=32$ using the special memory bypass mode (section \ref{sec:addrseq}), except that the technique described below is used to improve performance for $N \ge 1024$.  Fig.\ \ref{fig:performance} also shows the chip input and output rates, making it apparent when performance is limited by the maximum input rate of 16 Gb/s or output rate of 8 Gb/s (500 Mb/s per I/O pin).

\subsection{Large $N$}\label{sec:largeN}

Although we could continue the extrapolation of Table \ref{tbl:pwrVsN} to larger $N$, performance would deteriorate rapidly.  Since the on-chip memory is finite, larger $N$ means that $T$ must be reduced in order to store $2NT$ samples.  Shorter integrations increase the output rate, which is $f_c n^2/T$ products per unit time, so $f_c$ must be reduced to stay within the available output rate capacity.  The result for our parameters of $n=64$ and $M=65536$ words is that $f_c$ and hence the overall processing rate is maximized at $N=640$.  Below this, the device is input-rate limited, and above this it is output-rate limited.  This is seen in Fig.\ \ref{fig:performance}.  However, at large $N$ there is a way to do better.  This is exploited in the results of Fig.\ \ref{fig:performance}, which shows that $f_c$ and $b$ are 1.2 times larger than in Table \ref{tbl:pwrVsN} at $N=1024$ and 2.7 times larger at $N=2048$.  This enables maintaining nearly the same integration length $T$, processing rate $f_c$, and output rate at arbitrarily large $N$.  Otherwise, the processed bandwidth per device would be proportional to $N^{-3}$ rather than $N^{-2}$.

\begin{figure}[h]
%\vskip -0.10in
\begin{center}
\includegraphics[height=1.2 in]{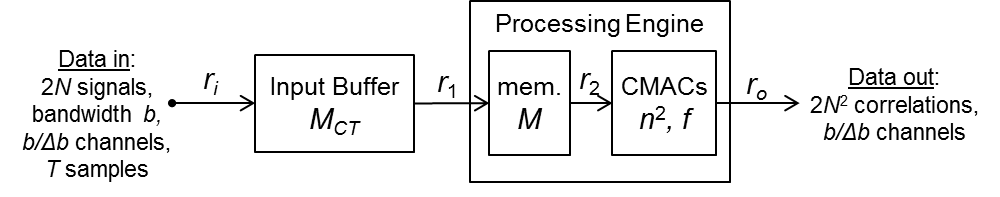}
\end{center}
\vskip -0.15in
\caption{Correlation unit with input buffer.}
\label{fig:extmem}
\end{figure}

\begin{figure}[h]
%\vskip -0.10in
\begin{center}
\includegraphics[height=2.5 in]{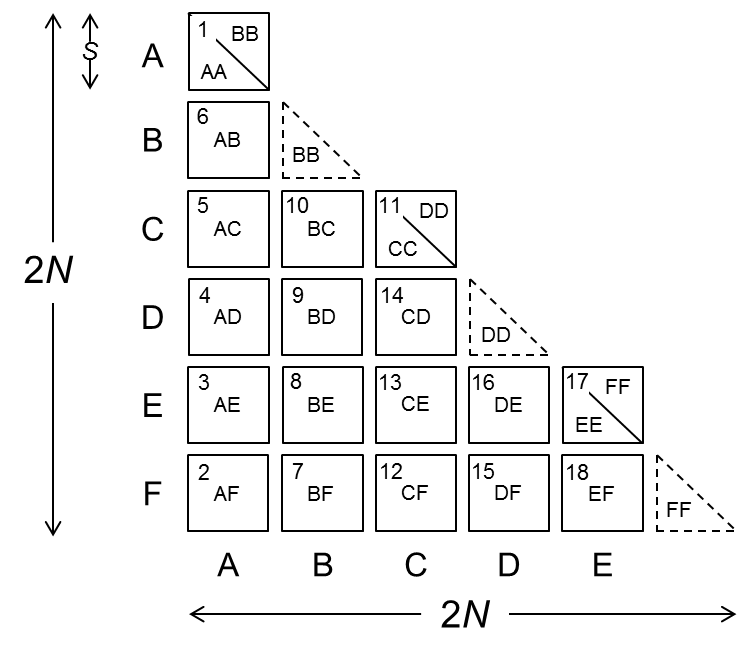}
\end{center}
\vskip -0.15in
\caption{Partial integration (PI) arrangement when $N/S = 3$.  Each square represents one PI. 
The $2N$ signals are partitioned into six groups of $S$ signals, A through F,
and two groups are correlated in each PI. On the diagonal of the matrix, a single PI is able
to process all self- and cross-correlatons within each of two groups. The dashed
triangles represent processing that was done earlier.
The numbers in the upper left of each square give the optimum sequence of PIs
such that only one group is different from those of the previous PI, minimizing the input 
data rate.  Such a sequence exists for any integer value of $N/S$.}
\label{fig:PI}
\end{figure}

The method involves using a second level of buffering and loading only part of the $2NT$ samples into the processing chip at a time, as shown in Figure \ref{fig:extmem}.  
The input buffer's size $M_{CT}$ must be sufficient to hold $2NT$ samples, but we now partition the data into $2N/S$ subsets of samples from $S$ of the signals, where $N/S$ is an integer.  Then $2ST$ samples from two of the subsets are loaded into the processing chip.  The processing of these signals completes a {\it partial integration} (PI).  Additional size-$S$ subsets are then loaded and additional PIs are computed in such a way that all correlations among the $2N$ signals are finally computed, as illustrated in Figure \ref{fig:PI}.  In this way, integration length $T$ can be nearly a factor of $N/S$ larger than it could be if we had to load all $2NT$ samples at once (full-integration processing).  This allows the output data rate to be lower, avoiding the output rate limit.  It comes at a cost of increasing the input rate by the same factor, since each subset must be loaded $N/S$ times.  Processing via PIs is thus useful only at large $N$, where full-integration processing would be output-rate limited and at least a factor of two below the input rate limit.  For our design, this happens at $N \ge 1024$.

A complete integration requires $2(N/S)^2$ partial integrations.  Each PI processes two different sets of $S$ signals, usually computing the $S^2$ cross-correlations of the first set with the second set, but sometimes computing the correlations among the signals of the first set and also those among the signals of the second set (see Fig.\ \ref{fig:PI}).  Just as with full integrations, a PI is broken into sub-integrations within the processing chip.  A total of $(S/n)^2$ SIs is needed to process either type of PI, or $2(N/S)^2\,(S/n)^2 = 2(N/n)^2$ SIs altogether, exactly the same as would be needed if all signals were processed at once.  The CMAC re-use factor is unchanged, but throughput is higher because the CMACs can run faster without exceeding the ouput rate limit.

To allow PIs to be processed continuously, one set of $ST$ samples must fit in memory space $M/3$, so that both sets of the current PI fit in $2M/3$.  The remaining $M/3$ is then used to receive the $ST$ samples that will be needed for the next PI.  This means that the overlap memory is 50\% of that holding the data being processed, vs.\ a maximum of 33\% for full-integration processing.  This limits the increase in $T$ for PI processing to a factor of $8N/9S$ rather than $N/S$.  The memory address sequence needed to process all SIs of one PI is also somewhat different from that shown in Figs.\ \ref{fig:subint} and \ref{fig:addrplot}, but the address sequence within an SI is the same, so no change to the chip's logic is needed \cite{lrd2015b}.

The input buffer memory must be external to the processing chip, but such a memory is also needed for full-integration processing because the data needs to be re-ordered, as explained in section \ref{sec:integration}.  The required capacity of the memory is the same for partial-integration and full-integration processing.  The difference is that with PI processing each sample is read $S/N$ times, so the buffer's output rate  is higher than its input rate by that factor, but it is never higher than the input bandwith of the processing chip, which is 16 Gb/s for our design.  The higher data rate increases the power consumption of the external devices and thus might contribute to a worsening the system-level energy FoM.  This is compensated by improved chip-level FoM, but more importantly by a reduction in the static (leakage) power of all devices since fewer X-units are needed to process a given total bandwidth.  Whether the overall power consumption is higher or lower with PI processing depends on the parameters of a particular application and the efficiency of the external devices, but the cost will certainly be lower.  For a given (large) $N$, lower system power consumption could be obtained by using a larger processing chip with more memory, but the use of PIs enables a chip with fixed capacity to be used efficiently at arbitrarily large $N$.

\section{Board and System Integration}\label{sec:integration}

In a typical application, multiple correlation ICs are used to implement part of an FX correlator, as shown in Figure \ref{fig:arch}.  The ICs are used in the ``X" portion, and each is the main component of one ``X unit."  The complete correlator also includes an ``F" or frequency analysis portion and a system for interconnecting the F and X portions.  If the system bandwidth $B$ is greater than the bandwidth $b$ that one chip can correlate, then $K = B/b$ chips (and X units) are needed.

\begin{figure}[h]
%\vskip -0.10in
\begin{center}
\includegraphics[height=2.5 in]{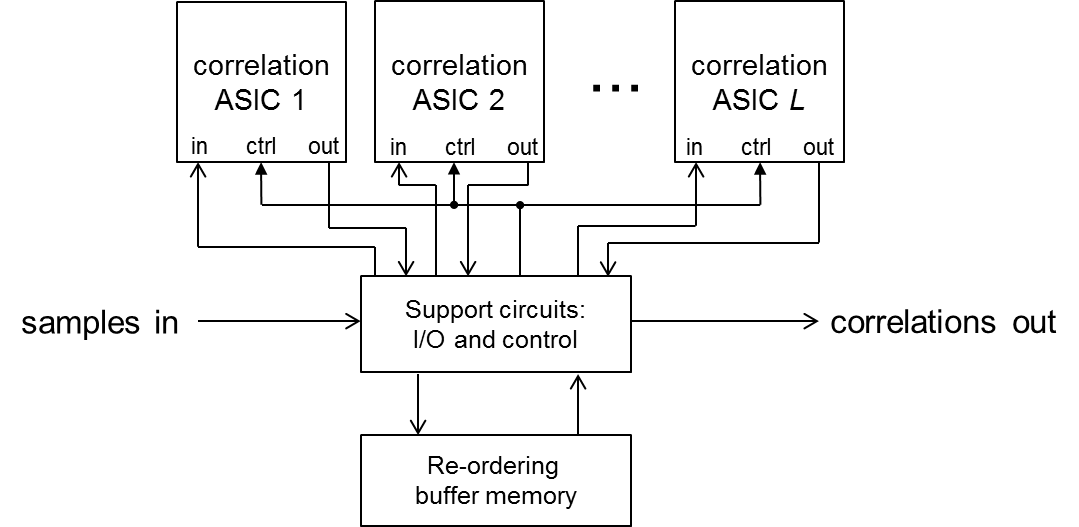}
\end{center}
\vskip -0.15in
\caption{Board-level design concept.  Each board provides multiple correlation units, each using one of ICs of this paper.  Supporting logic common to all correlation units is needed, including board-level I/O and a data reordering buffer.}
\label{fig:board}
\end{figure}

Normally one or more of the ICs is installed on a printed circuit board and integrated with supporting logic, and multiple similar boards are used to build the complete X portion of the correlator.  A general block diagram for one such board is shown in Figure \ref{fig:board}.  
The supporting logic must provide the board-level input and output interfaces and generate the necessary clocks and control signals.  The control signals include INTEGRATE and the memory address sequence that is written to each chip's registers via the SPI port.  The board must also provide an input data buffer for putting the samples into the order required by the correlation chips.  In many situations, the supporting logic can be provided by an FPGA and one FPGA can support multiple correlation ICs.

It is necessary for the F portion to implement a filter bank that breaks each signal into channels of width no larger than $b$, and an interconnection network is needed to send the samples for different channels to different correlation chips.  In practice it is often desired to have channels of width $b_c < b$, in which case each IC must process $c = b/b_c$ channels.  This is possible because the sampling rate for each channel is only $b/c$, so the IC can process it in a time smaller by factor $b_c/b$ than the time over which those samples were acquired.  It can therefore keep up with the incoming data by processing the channels sequentially.

The re-ordering buffer is needed because the natural order of the F section outputs is different from that required at the X unit input.  Although the specific order needed here is peculiar to the design of our chip, a buffer of the same size would be needed for any FX correlator.    The natural order of incoming data has one sample from all $2N$ signals followed by the next sample from all signals, etc., whereas for cross-correlation we need $T$ samples for one group of signals (64 in our case) followed by $T$ samples for the next group, etc.  The buffer must hold $2NTc$ samples, enough for one integration; in order for the data to be written and read in different orders it must be doubled to $4NTc$ samples.  It can be external to the processing chip because (for full integrations) its writing and reading rates are the same and each sample is written and read exactly once.  With partial integrations (section \ref{sec:largeN}), the reading rate must be higher, but PI mode is used only at large $N$ where the input rate per X unit is lower.

\subsection{Design Example}

Consider an array of $N = 512$ dual-polarization antennas producing signals of bandwidth $B = 1000\,$MHz.  From Table \ref{tbl:pwrVsN}, the processing chip can operate at a CMAC rate of $f_c= 250\,$MHz and requires a re-use factor of $x=128$, so each chip can process a bandwidth of $b = f_c/x=1.95\,$MHz.    We therefore need $B/b = 512$ chips to correlate the full bandwidth.  The chip's input and output rates are 16 Gb/s and 4.72 Gb/s, respectively.  The memory is large enough for $T = 6944$, so a re-ordering buffer of at least $4NT= 14,221,312$ samples or 108.5\,Mib (at $w_i=8$\,b per sample) will be needed for each chip.  If the signals are broken into 61\,kHz channels, then each IC must process $c = 32$ channels and the re-ordering buffer size becomes $4NTcw_i = 3.39$\,Gib per chip.
  
\begin{figure}[h]
%\vskip -0.10in
\begin{center}
\includegraphics[height=3 in]{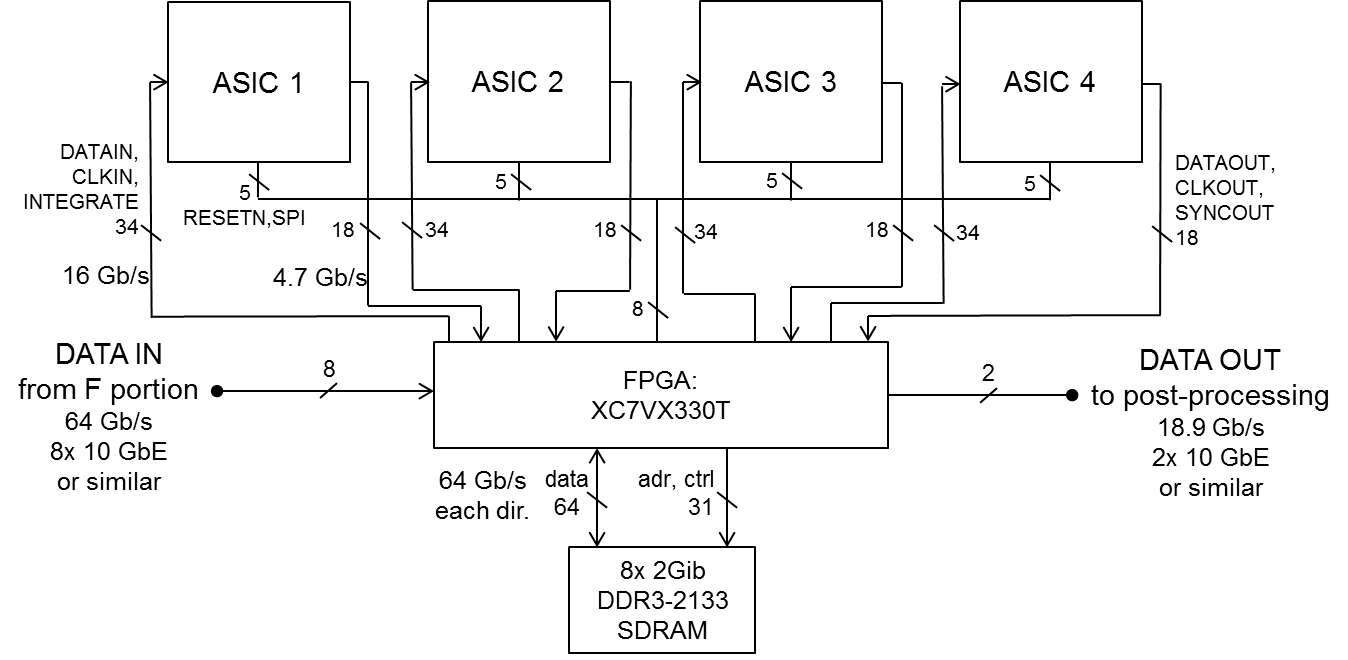}
\end{center}
\vskip -0.15in
\caption{Sample board-level design of a correlator for $N=512$.  Four of our ICs are supported by one FPGA and a set of commodity memory chips.  Each correlation IC dissipates 1.82\,W and the total dissipation of the board is 16.5\,W.  To achieve a total bandwidth of $B=1000\,$MHz, 128 of these boards are needed.}
\label{fig:example}
\end{figure}

At the board level, at least four  correlation ICs can be supported by one currently-available FPGA (e.g., Xilinx Virtex 7, XC7VX330T) and one set of DDR3-2133 memory chips, as shown in Figure \ref{fig:example}.
The FPGA provides 8 high-speed serial ports (8 Gb/s each) for receiving the input data for all the correlation ICs, and 2 similar ports for transmitting the output data.   The data could be organized into 10 Gb/s Ethernet frames or into any of various other standard protocols.  

From Table \ref{tbl:pwrVsN}, each IC's power dissipation is 1.82\,W.  The FPGA is estimated to dissipate 5.55\,W \cite{xilinx}, and the memory chips (8 at 8x256K) 3.7\,W \cite{micron}, giving 16.5\,W of dissipation per board for signal-processing devices.  Allowing 20\% for power supplies and other overhead then gives 19.8\,W per board.  The system requires 128 boards, resulting in a total X-portion dissipation of 2.54\,kW.  The system-level energy figure of merit is then 4.84\,pJ per CMAC operation.

This example uses 28-nm FPGAs and DDR3 memory devices that are readily available now.   More advanced devices of both types are becoming available [including 20\,nm and eventually 14\,nm FPGAs \cite{altera14nm}, and Hybrid Memory Cubes \cite{hmc}].  Use of these would decrease the power for supporting devices and allow more correlation ICs per board, resulting in a system-level power efficiency approaching that of the correlation chips.

\section{Conclusions}

The IC design presented here provides a power-efficient means of computing all cross-correlations among many signals.  The power efficiency is more than two orders of magnitude better than that of existing large correlators, and about a factor of 10 better than planned correlators based on future-generation FPGAs.  The IC is flexible, in that it can be used to construct correlators for almost any number of antennas and any bandwidth, although its efficiency is best if $N$ is a multiple of 64.

The design has been synthesized and subjected to careful post-synthesis simulation and analysis.  From this we know that it will be smaller than 12 mm$^2$ when fabricated in the IBM 32SOI process, and we have accurate estimates of its power consumption.  However, the physical design (detailed layout) has not yet been done and devices have not yet been fabricated.

In the radio astronomy community, there is currently a strong preference for building correlators from off-the-shelf programmable components like FPGAs and GPUs.  So far, those devices have been used only for relatively small correlators.  Technology advances through Moore's Law continue to increase the size of digital processing machines that can be built that way, but custom-designed devices also benefit from these advances.  The argument is sometimes made that when faced with requirements that cannot be reasonably met by today's off-the-shelf devices, it is better to wait a few years until the devices get better than to spend that time developing a custom device.  The results presented here belie that argument.  We can do much better with devices fabricated in a technology that has been available for 5 years (32 nm) than with FPGAs that are two generations more advanced (14-16 nm) and not yet available.

The development cost of an ASIC is sometimes a deterrent to its adoption.  For the large telescopes now planned or underway, it is a negligible fraction of the total development cost and is recoverable by energy cost savings in the first few years of operation.  For smaller projects, several may be able to share the up-front costs provided that the design is flexible enough to be used in all of them.  For our design, a substantial part of the development work has already been done.  

The design presented here would be a good choice for telescopes whose designs will be frozen in the next few years.  For correlators needed further in the future, it would be reasonable to port the design to a more advanced technology, such as the 14 nm finFET processes that are now becoming available.  In such a next-generation design, other improvements could be considered, such as adding high-speed serial I/O (currently feasible but cost-prohibitive) and increasing the sizes of the CMAC array and memory.

\section*{Acknowledgments}

% Language required by R&TD rules
This work was carried out at the Jet Propulsion Laboratory, California Institute of Technology, under a contract with the National Aeronautics and Space Administration and funded through the internal Research and Technology Development program.

We thank Dayton Jones, Joseph Lazio and Sander Weinreb of JPL/Caltech for advice and encouragement throught this project.  Jim Maryoung of Synopsys Inc.\ provided an exceptional level of assistance in the use of his company's tools.   David Hawkins, Jonathon Kocz, and Sander Weinreb provided valuable comments on an early draft.

%(Additonal acknowledgments to be written)

\section*{Appendix A:  Clock Tree Synthesis}

Here we describe a manual, non-optimum synthesis of the clock trees for the purpose of estimating their power and area.   Final design of clock trees will be done during layout.

Two library cells were selected for building the trees, and their main properties are given in Table \ref{tbl:bufcells}.  Both are from the high-threshold-voltage library.
% Table bufcells goes here.
\begin{wstable}[h]
\caption{Library Data on Standard Cells Used for Clock Tree Buffers}
\begin{tabular}{@{}lrrrrrr@{}} 
\toprule
cell name  & input cap. & intrinsic delay & delay/load& leakage pwr & internal energy & area \\
               &\hfill fF\hfill\hfill  &\hfill ps\hfill\hfill &\hfill ps/fF\hfill\hfill &\hfill $\mu$W\hfill\hfill &\hfill fJ\hfill\hfill & $\mu\rm{m}^2$\\
\colrule
BUF\_X16M & 3.80112 & 14.61741 & 0.17408 & 0.022023 & 10.410215 & 2.691 \\
INV\_X9M & 6.55092 & 2.71712 & 0.58739 & 0.009357 & 2.49423 & 1.287 \\
\botrule
\end{tabular}
\label{tbl:bufcells}
\end{wstable}
We adopt the constraint that the delay through the tree should be less than 20\% of the clock period, and we use the maximum clock rate of each net (500 MHz for INCLK and 360 MHz for \sysclk\ and CLK\_CMAC).  This gives maximum delays of 400 ps for INCLK and 555 ps for \sysclk\ and CLK\_CMAC.

%A simple analysis shows how many levels of buffering are needed.  Let the fanout of every buffer in the tree be the same, $F$.  Let the root of the tree contain a single buffer that drives $F$ buffers at level 1, each of which drives $F$ at level 2, continuing until each buffer at the final level drives $F$ destination loads.   If there are $L$ destination loads, then $K = \ceil(L/(F+1))$ levels and $1 + F + F^2 + ... + F^L$ buffer cells are needed.  From this it is easily shown that $F \ge 24$ gives $K \le 3$ for $L =$311,488 (CLK\_CMAC net) and $K \le 2$ for $N = $8,352 or less (\sysclk\ and INCLK).  Since there are $K + 1$ levels of nets, let the maximum delay per level be $1/(K+1)$ of the maximum delay for the tree; this gives worst-case delays per level of 133 ps for INCLK, 185 ps for \sysclk\ and CLK\_CMAC, all for $K = 2$.

To determine the delay, we adopt the rule-of-thumb that the wire load on a net is 2 fF for each destination cell input\footnote
{This is justified by experience with this design.  Neglecting I/O cells and clock trees, the total switching power with Design Compiler Topographic's estimate of the wire load is 152.3\,mW (Table \ref{tbl:pcresults}).  The corresponding value with no wire load, using the synthesized netlist and delays from a compiler run with ideal wires (no capacitance and no delay) is 78.4\,mW.  There are other differences between these synthesized designs, but this indicates that, on average, the total load is about twice the destination pin load.  Since the average input pin load throughout the design is about 1.9\,fF, we adopt a conservative estimate of 2\,fF of wire load per destination cell.}%
, and we neglect the network delay because the chip is small (about 3.4 mm edge-to-edge, giving a network delay of about 19 ps at a dielectric constant of 2.7).  The selection of the two standard cells in Table \ref{tbl:bufcells} was made by considering various pairs of possible buffer cells and finding the maximum number of one that can be driven by the other while keeping the delay sufficiently low.  From this we find that three levels of buffering are needed for the CLK\_CMAC network (311,488 loads), but two are sufficient for \sysclk\ and INCLK.  The buffer trees constructed from these cells will certainly not be optimum, but we are not attempting to produce the optimum design in this simple exercise; we merely want to produce a feasible design and evaluate its area and power.  It will then be apparent whether further optimization is worthwhile.

Since these are all clock nets, the final destination loads are the clock pins of flip flops.  There are 6 types of D FF in the library, and each has versions with 3 or 4 drive strengths, but among all of them the clock pin capacitance ranges from 0.499 pF to 0.672 pF.  The maximum capacitance was used for our delay calculations.

From the above considerations, we choose the following rough designs:\hfill\break
\vbox{
\smallskip
\begin{tabular}{rll}
\multicolumn{3}{l}{CLK\_CMAC, 311488 loads:} \\
\qquad 2686x & INV\_X9M   & fanout 116, driving the destination loads \\
\qquad    75x & INV\_X9M    & fanout 36, driving above buffers \\
\qquad      1x & BUF\_X16M  & fanout 75, driving above buffers. \\
\multicolumn{3}{l}{\sysclk, 8352 loads:}\\
\qquad     48x  & INV\_X9M & fanout 175, driving the destination loads \\
\qquad       1x  & INV\_X9M & fanout 48, driving the above buffers. \\
\multicolumn{3}{l}{INCLK, 3168 loads:}\\
\qquad     26x & INV\_X9M & fanout 125, driving the destination loads \\
\qquad       1x & INV\_X9M & fanout 26, driving the above buffers. \\
\end{tabular}
\smallskip
}\hfill\break
Using these designs and the data of Table \ref{tbl:bufcells}, Table \ref{tbl:bufresults} shows the results of calculating the area, power, and delay of the three buffer trees.  The frequency of each clock is the effective frequency in the scenario that was simulated in Section \ref{sec:sim}, not the maximum used for timing.  The switching power of nets driving the destination cells was already included in the Power Compiler results (Table \ref{tbl:pcresults}), so it is excluded from the ``added power'' in Table \ref{tbl:bufresults}.  The total power added by these buffer trees is then about 39.65 mW.  Optimized buffer trees will use less.  The added cell area is 3654 $\mu\rm{m}^2$.  The calculated delay through each buffer tree is also given, showing that each is less than the limit we specified.
\begin{wstable}[h]
\caption{Calculation of Power and Area for Clock Buffer Trees}
\begin{tabular}{@{}lrrrrrrrr@{}} 
\toprule
Net &\multicolumn{3}{c}{CLK\_CMAC} & \multicolumn{2}{c}{sysclk} & \multicolumn{2}{c}{INCLK} & Totals \\
Max.\ freq.\tnote{a}, MHz &\multicolumn{3}{c}{360}  & \multicolumn{2}{c}{360}      & \multicolumn{2}{c}{500}& \\
Frequency\tnote{b}, MHz  &\multicolumn{3}{c}{62.5} & \multicolumn{2}{c}{227.273}& \multicolumn{2}{c}{500}& \\[-1\medskipamount]
& \multicolumn{3}{c}{\hrulefill} & \multicolumn{2}{c}{\hrulefill} & \multicolumn{2}{c}{\hrulefill} \\
%\cline{2-4} \cline{5-6} \cline{7-8} \\
Cell&INV\_X9M&INV\_X9M&BUF\_X16M&INV\_X9M&INV\_X9M&INV\_X9M&INV\_X9M& \\
Number&2686&75&1&48&1&26&1& \\
Fanout per cell&116&36&75&175&48&125&26& \\
Load per cell, fF&309.953&307.833&641.319&467.601&410.444&334.001&222.324& \\
Leakage, $\mu$W&25.13&0.70&0.02&0.45&0.01&0.24&0.01&26.57 \\
Internal, $\mu$W&837.44&23.38&1.30&54.42&1.13&64.85&2.49&985.02 \\
Switching, $\mu$W&42146.98&1168.80&32.47&4131.90&75.56&3517.03&90.04&51162.78 \\
Delay, ps&184.7793&183.5343&126.2569&277.3799&243.8067&198.9048&133.3073& \\[-1\medskipamount]
& \multicolumn{3}{c}{\hrulefill} & \multicolumn{2}{c}{\hrulefill} & \multicolumn{2}{c}{\hrulefill} \\
%\cline{2-4} \cline{5-6} \cline{7-8} \\
Total power, $\mu$W& 
                        \multicolumn{3}{c}{44236.23}& \multicolumn{2}{c}{4263.47} & \multicolumn{2}{c}{3674.67}&52174.36 \\
Load pin sw., $\mu$W&
                         \multicolumn{3}{c}{10599.91}&\multicolumn{2}{c}{1039.17} & \multicolumn{2}{c}{884.53} & \\
Added power, $\mu$W&
                         \multicolumn{3}{c}{33636.32}& \multicolumn{2}{c}{3224.30}& \multicolumn{2}{c}{2790.14}&39650.76 \\
Total delay, ps & \multicolumn{3}{c}{494.5705}& \multicolumn{2}{c}{521.1866}& \multicolumn{2}{c}{332.2121}& \\
Total cell area, $\mu\rm m^2$ &
                        \multicolumn{3}{c}{3556.0980}&\multicolumn{2}{c}{63.0630}& \multicolumn{2}{c}{34.7490}& 3653.9100\\
\botrule
\end{tabular}
\begin{tablenotes}
\item[a] Maximum switching frequency, for delay constraint.
\item[b] Effective switching frequency in the simulated scenario, for power calculation.
\end{tablenotes}
\label{tbl:bufresults}
\end{wstable}

%\bigskip
%\parindent=0pt
%\obeylines{
%More References --- to be filled in and put into BibTex format
%[f] SPI
%[3] something about IBM 32SOI.  {32soi}
% http://www.eetimes.com/document.asp?doc\_id=1254655
% https://www-03.ibm.com/press/us/en/pressrelease/28428.wss
% http://www.edn.com/electronics-blogs/professor-memory/4413919/IBM-s-Cu-32-Custom-Logic-SOI-Process-with-Embedded-DRAM-Memory
%[4] Synopsys {synopsys}
%[5] ARM standard cell library {arm}
%[6] Modelsim {modelsim}
%[7] Memo No.~14 (for SPICE simulations of output) {lrd2015c}
%[7a] SPICE {spice}
%[8] Alterna Stratix V data sheet. {altera}
%[9] SDF file definition {sdf}
%[10] SAIF file definition  (citation replaced by footnote)
%[12] Flip chip packaging {flipchip}
%[13] Xilinx power estimator {xlinix}
%[14] DDR3 data sheet {micron}
%[15] 20nm and 14nm FPGAs
%[16] HMCs
%}

\end{document}